\documentclass[]{JHEP3}
\usepackage{epsfig}
\usepackage{amsmath,amsfonts,amssymb}

\newcommand{\dis}[1]{\begin{equation}\begin{split}#1\end{split}\end{equation}}

\newcommand{\be}{\begin{equation}}

\newcommand{\Mp}{M_P}

\newcommand{\tev}{\,\textrm{TeV}}
\newcommand{\gev}{\,\textrm{GeV}}
\newcommand{\mev}{\,\textrm{MeV}}

\newcommand{\gravitino}{{\widetilde{G}}}
\newcommand{\mgravitino}{{m_{\widetilde{G}}}}

\newcommand{\photino}{\widetilde{\gamma}}
\newcommand{\zino}{\widetilde{Z}}
\newcommand{\wino}{\widetilde{W}}
\newcommand{\sneutrino}{\tilde{\nu}}

\newcommand{\mz}{{M_Z}}
\newcommand{\mw}{{M_W}}
\newcommand{\cv}{{C_V}}
\newcommand{\ca}{{C_A}}
\newcommand{\Uz}{{U_{\zino\zino}}}
\newcommand{\Uw}{{U_{\wino\wino}}}

\title{Indirect detection of gravitino dark matter including its three-body decays}
\author{Ki-Young Choi\\Department of Physics, Pusan National University, Busan 609-735, Korea\\ Email: \email{kiyoung.choi@pusan.ac.kr}}
\author{Diego Restrepo\\ Instituto de Fisica, Universidad de Antioquia, A.A.1226, Medellin, Colombia\\Email: \email{restrepo@udea.edu.co}}
\author{Carlos E. Yaguna\\Departamento de Fisica Teorica and Instituto de Fisica Teorica UAM-CSIC, 
Universidad Autonoma de Madrid, Cantoblanco, E-28049 Madrid, Spain\\Email: \email{carlos.yaguna@uam.es}}
\author{Oscar Zapata\\Escuela de Ingenieria de Antioquia, A.A.7516, Medellin, Colombia and Universidad de Antioquia, A.A.1226, Medellin, Colombia\\Email:\email{pfozapata@eia.edu.co}}
\abstract{It was recently pointed out that in supersymmetric scenarios with gravitino dark matter and bilinear R-parity violation, gravitinos with masses below $M_W$ typically decay with a sizable branching ratio into the 3-body final states $W^*\ell$ and $Z^*\nu$. In this paper we study the indirect detection signatures of gravitino dark matter including such final states. First, we obtain the  gamma ray spectrum from gravitino decays, which features a monochromatic contribution from the decay into $\gamma\nu$ and a continuum contribution from the three-body decays. After studying its dependence on supersymmetric parameters, we compute the expected gamma ray fluxes and  derive new constraints, from recent FERMI data, on the R-parity breaking parameter and on the gravitino lifetime. Indirect detection via antimatter searches, a new possibility brought about by the three-body final states, is also analyzed. For models compatible with the gamma ray observations, the positron signal is found to be negligible whereas the antiproton one can be significant.}

\preprint{PNUTP-10-A08}
\begin{document}

\section{Introduction}
Supersymmetric scenarios are considered to be the most promising extensions of the Standard Model. They are motivated by the hierarchy problem, by the unification of the gauge couplings, and by the existence of dark matter. Recent observations indicate that dark matter accounts for about $25\%$ of the energy-density of the Universe \cite{Dunkley:2008ie}, a fact that cannot be explained within the standard model of particle physics.  Supersymmetric models, on the other hand, contain at least two viable dark matter candidates: the neutralino and the gravitino. Both  give rise to an interesting phenomenology and have been extensively studied in the literature. In this paper we focus on models with gravitino dark matter.

In supersymmetric models, R-parity conservation was long believed to be a
prerequisite for supersymmetric dark matter, for it guarantees the  stability
of the lightest supersymmetric particle (LSP) --the would-be dark matter
candidate.  It was pointed out in \cite{Takayama:2000uz}, however, that if  the gravitino --the superpartner of the graviton that  arises in local
supersymmetric theories-- is the LSP, it  can be a suitable dark matter candidate  in such scenarios.
 Indeed, even though the gravitino is unstable due to R-parity breaking, 
 its lifetime may be much longer than the age of the Universe. Such a long lifetime is the result of the gravitino weak interactions, which are suppressed by the Planck scale and by  small R-parity violating couplings. A scenario with gravitino dark matter and R-parity violation is also favoured by  thermal leptogenesis, as it alleviates the tension between the large reheating temperature required by leptogenesis and the constraints from  Big-Bang Nucleosynthesis \cite{Buchmuller:2007ui}. In R-parity violating models, therefore, the primordial gravitino produced in the early Universe is  a  viable and well-motivated dark matter candidate.

A salient feature of such  models is that, unlike their R-parity conserving counterparts, gravitino dark matter can be indirectly detected. In fact, the small fraction of gravitinos that have decayed until today constitutes a source of high energy cosmic rays \cite{Bertone:2007aw,Ishiwata:2008cu} that could be observed in present and future experiments. Gravitinos may be indirectly detected via  gamma rays \cite{Bertone:2007aw,Ibarra:2007wg}, neutrinos \cite{Covi:2008jy} or  antimatter searches \cite{Ibarra:2008qg}. Present data from FERMI, for instance, already constrains the parameter space of these models \cite{Choi:2009ng}.

The indirect detection signatures of gravitino dark matter  strongly depend on the gravitino lifetime and on its branching ratios. It was recently demonstrated that, in bilinear R-parity breaking models, gravitinos with masses below $M_W$ not only decay into $\gamma\nu$, as previously believed, but also into the three-body final states $W^*\ell$ ($\to \sum_f f\bar f'\ell$) and $Z^*\nu$ ($\to \sum_ff\bar f \nu$) \cite{Choi:2010xn}. Because the branching ratio into these three-body final states is typically sizable, they considerably alter the gravitino lifetime and the indirect detection signatures of gravitino dark matter. In this paper we revisit the prospects for the indirect detection of gravitino dark matter in view of these new decay modes of the gravitino. 

In the next section we introduce the model and explain the main assumptions of our work.  Then, we briefly revisited the dominant decay modes  of gravitino dark matter. In section \ref{sec:gammas}, the expected gamma ray flux from gravitino decay is studied, in particular its dependence on gaugino and gravitino masses. Then, we use  that flux and recent Fermi data to derive constraints on the R-parity breaking parameter and on the gravitino lifetime. In section \ref{sec:anti},  the indirect detection of gravitino dark matter via antimatter searches is considered. After deriving the spectra of positrons and antiprotons, we obtain the expected fluxes at earth and compare them with present data. For models compatible with the gamma ray observations, the positron signal is predicted to be negligible whereas the antiproton one can be significant.
\section{Bilinear R-parity violating models}
We work in the framework of  supersymmetric models with bilinear R-parity
violation and gravitino dark matter, such as those considered in~\cite{Buchmuller:2007ui,Ishiwata:2008cu,Ibarra:2007wg}. That is, we assume that the gravitino is the lightest supersymmetric particle and that it accounts for the observed dark matter density of the Universe, $\Omega_{\gravitino}h^2\approx 0.11$. In these models, the gravitino is unstable, due to R-parity breaking, and may decay into standard model particles via lepton number violating interactions.

In bilinear R-parity violating models, the superpotential is given by
\begin{equation}
W=W_{\text{MSSM}}+\mu_iL_iH_u,
\label{eq:sp}
\end{equation}
where $W_{\text{MSSM}}$ is the usual (R-parity conserving) superpotential of the MSSM, $L_i$ ($i=e,\mu,\tau$) is the lepton doublet, $H_u$ is the up-type Higgs doublet, and $\mu_i$ are R-parity violating couplings. To study these models, it is convenient to redefine the Higgs and lepton doublets so as to eliminate the bilinear R-parity violating interactions from the superpotential. In that basis, which is the one we consider in the following,  the mixing terms between the Higgsino and the lepton doublet are eliminated from the fermion mass matrix, and  the soft-breaking Lagrangian contains the following terms
\begin{equation}
\mathcal{L}_{\text{soft}}=\mathcal{L}_{\text{soft}}^{\text{MSSM}}+B_i\tilde L_iH_u+m^2_{\tilde L_iH_d}\tilde L_i H_d^*+\mathrm{h.c.}
\end{equation}
Here  $H_d$ is the down type Higgs boson doublet, and $B_i, m^2_{\tilde L_iH_d}$ are R-parity violating parameters. Even though the original superpotential, equation (\ref{eq:sp}), did not contain any trilinear R-parity violating terms, the redefinition of the fields that eliminates the bilinear ones induces $\lambda$ and $\lambda'$ terms in the superpotential. They are given by
\begin{equation}
W_{\text{RPV}}=\lambda_{ijk} L_kL_iE^c_j+\lambda'_{ijk}L_kQ_iD^c_j\,,
\end{equation}
where $L_i$ and $Q_i$ are the lepton and quark doublets, whereas $E_i^c$ and $D_i^c$ are the lepton and down-quark singlets. The couplings $\lambda$ and $\lambda'$ are related to the original couplings of the superpotential by
\begin{equation}
\lambda_{ijk}=\frac{\mu_k}{\mu}\lambda^e_{ij}\quad \mathrm{and}\quad \lambda'_{ijk}=\frac{\mu_k}{\mu}\lambda^d_{ij}\,,
\label{eq:trilinear}
\end{equation}
where $\mu$ is the usual MSSM parameter and $\lambda^e$ and $\lambda^d$ are respectively the charged lepton and down-quark Yukawa couplings. As we will see, these trilinear terms play essentially no role in our discussion.

The two  R-parity violating terms present in $\mathcal{L}_{\text{soft}}$ give rise to non-zero vacuum expectation values for the sneutrino fields.  They  can be written as
\begin{equation}
\langle\tilde\nu_i\rangle=\frac{B_i\sin\beta+m^2_{\tilde L_iH_d}\cos\beta}{m^2_{\tilde\nu_i}}v,
\end{equation} 
where $v=174~\gev$, $\tan\beta=\langle H_u^0\rangle/\langle H_d^0\rangle$, and $m_{\sneutrino_i}$ is the sneutrino mass. It is useful to define the dimensionless parameters $\xi_i$, $i=e,\mu,\tau$, as
\begin{equation}
\xi_{i}\equiv \frac{\langle\sneutrino_i\rangle}{v}.
\end{equation}
Because the R-parity violating couplings are expected to be largest for the third generation, we will assume, following~\cite{Covi:2008jy}, that the sneutrino acquires a vev only along the $\tilde\nu_\tau$ direction. In other words, we suppose that $\xi_\tau\gg \xi_\mu,\xi_e$. In this setup, all R-parity violating effects, including the decay of the  gravitino and non-zero neutrino masses, are controlled by $\xi_\tau$.

Since neutrino masses may arise from other sources, like the seesaw mechanism, we will only require that the contribution to neutrino masses from R-parity violation should not exceed the value obtained from present data. This condition implies that $\xi_\tau\lesssim 10^{-7}$. In addition, a lower bound on $\xi_\tau$  is obtained from successful big bang nucleosynthesis: $\xi_\tau\gtrsim 10^{-11}$ \cite{Ishiwata:2008cu}. A priori, therefore, $\xi_\tau$ can vary over four orders of magnitude. 
In section \ref{sec:gammas} we will use the gamma ray  flux from gravitino decays and recent Fermi data to derive new constraints on $\xi_\tau$.

\section{Gravitino decays}
In this section we briefly revisit the decays of the gravitino in bilinear R-parity violating scenarios. We assume throughout this paper that  $\mgravitino\lesssim M_W$, for it is only in such mass range that the three-body final states modify the gravitino lifetime and branching ratios.  The gravitino decay rates can be calculated directly from the interaction Lagrangian~\cite{Bagger:1990qh}. It turns out that gravitinos with masses below $M_W$ can decay into the two-body final state $\gamma\nu_\tau$ or into the three-body final states $W^*\tau$ and $Z^*\nu_\tau$. The decays induced by the trilinear couplings, equation (\ref{eq:trilinear}), are suppressed by the Yukawa couplings and by sfermion masses and turn out to be negligible compared to these (see e.g. \cite{Buchmuller:2007ui}), so we will not considered them in the following. For the two-body final states $\gamma\nu_\tau$, the gravitino decay width is 
given by~\cite{Ishiwata:2008cu,Covi:2008jy}
\begin{equation}
\Gamma(\gravitino\to \gamma \nu_\tau)=
\frac{\xi_\tau^2\mgravitino^3}{64\pi\Mp^2}|U_{\photino\zino}|^2,\\
\end{equation}
where $U_{\photino\zino}$ is the photino-zino mixing parameter \cite{Covi:2008jy}. To a good approximation it can be written as \cite{Ibarra:2007wg,Grefe:2008zz}
\begin{equation}
|U_{\photino\zino}|\approx \frac{M_Z(M_2-M_1)s_Wc_W}{(M_1c_w^2+M_2s_w^2)(M_1s_W^2+M_2c_W^2)},
\end{equation}
where $s_W$ and $c_W$ denote the sine and cosine of the weak mixing angle, and
$M_1,M_2$ are the $U(1)$ and $SU(2)$ gaugino masses. For simplicity, we will
assume in this paper that gaugino masses are universal at the GUT scale, so that $M_2\sim
1.9 M_1$ at the electroweak scale, and we will consider $M_1$ as a free parameter of the model. In any case, our results do not depend strongly on this universality assumption. Notice, from the above equations, that the mixing parameters and the decay rate into $\gamma\nu_\tau$ decrease with gaugino masses.
\vspace{5mm}

\begin{center}
\FIGURE{\begin{tabular}{ccc}\includegraphics[scale=0.34]{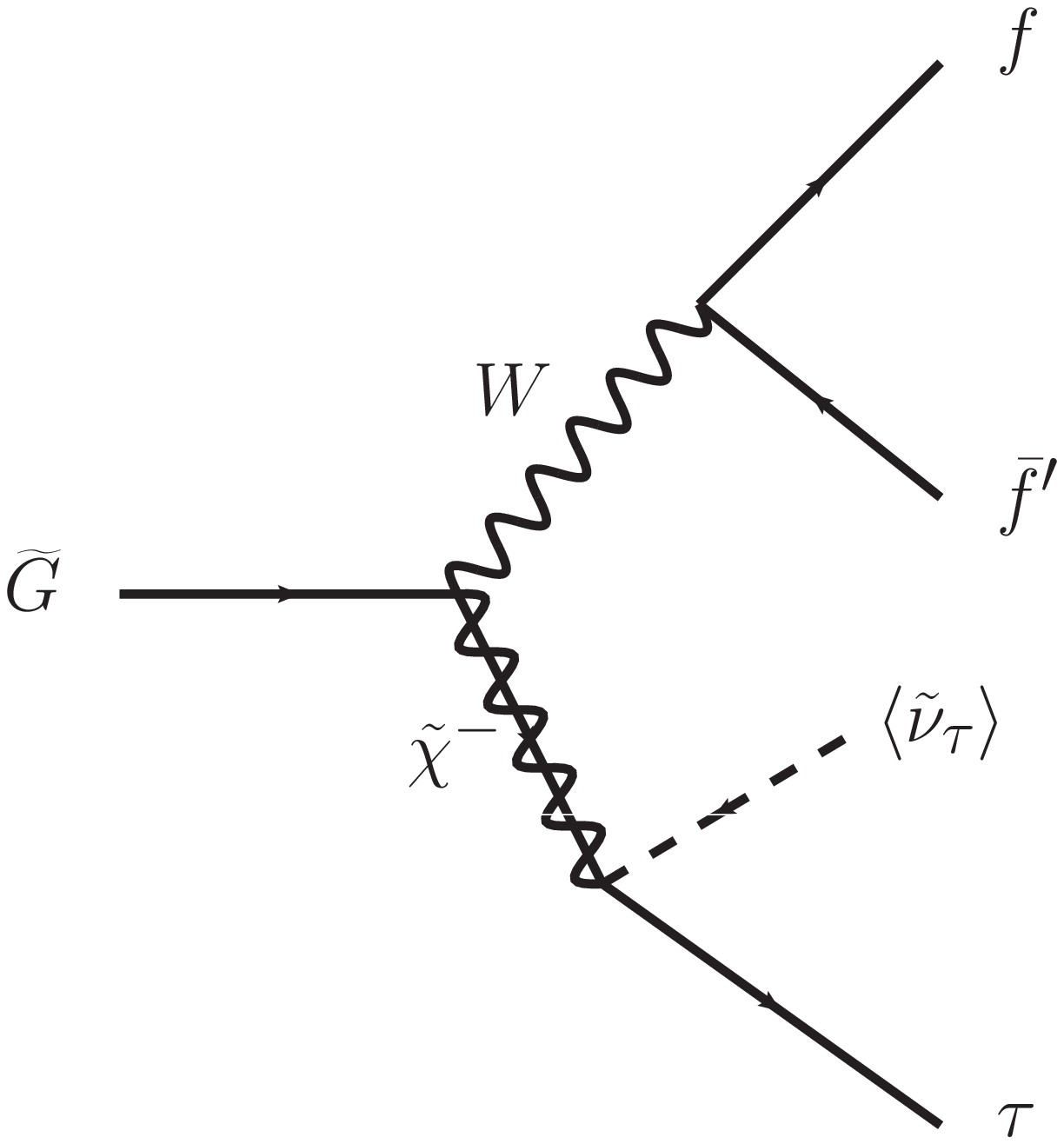} & \hspace{1cm} & \includegraphics[scale=0.34]{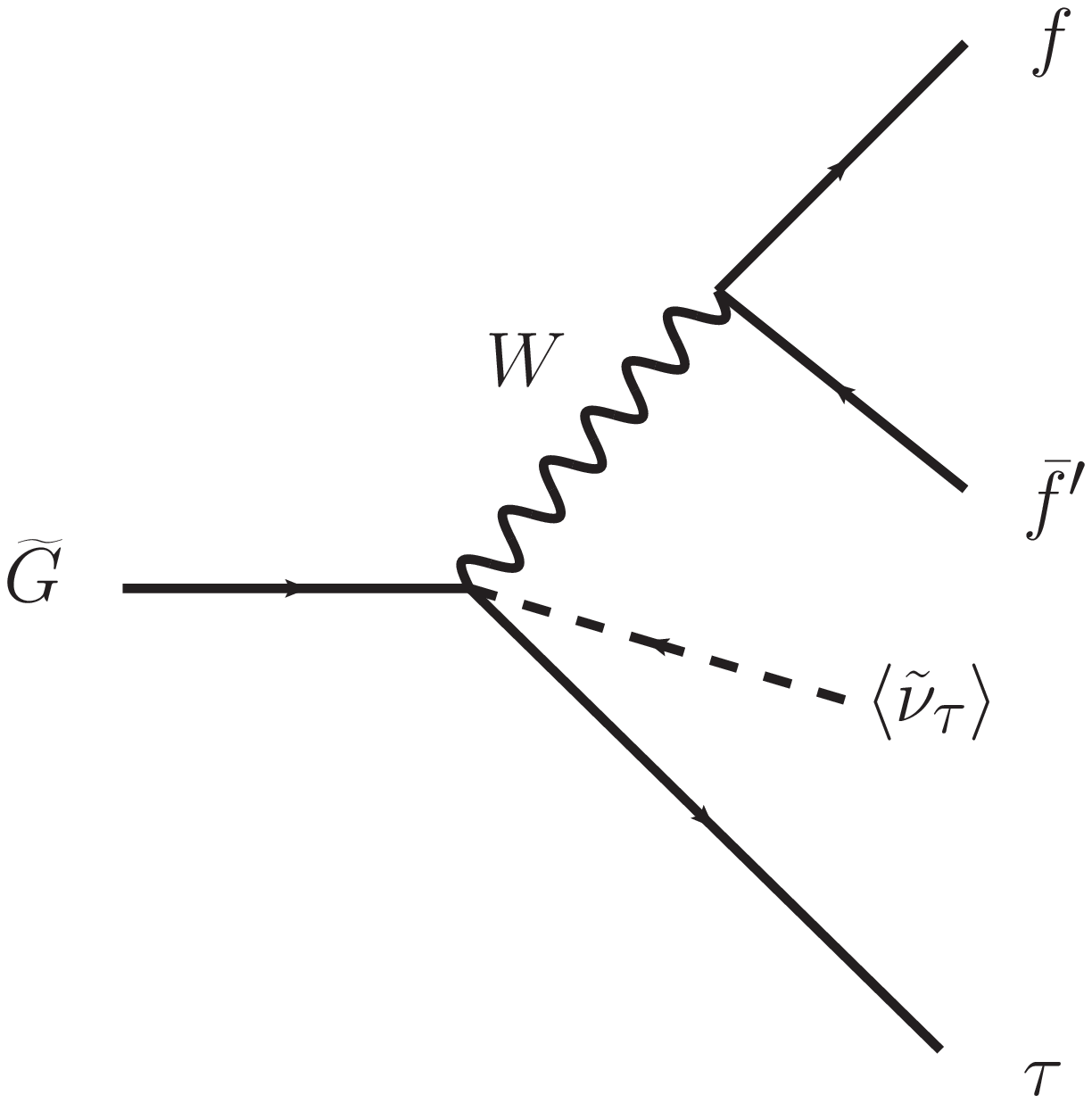}\end{tabular}\caption{The two diagrams that contribute to gravitino decay into $W^*\tau$}\label{fig:Wdiag}}
\end{center}

The  gravitino decay rates into the three-body final states $W^*\tau$ and $Z^*\nu_\tau$ were first computed in \cite{Choi:2010xn}. Each of these processes receives contributions from two different diagrams, as illustrated in figure \ref{fig:Wdiag} for the decay into $W^*\tau$. In the appendix we provide the analytical expressions for both decay rates. A crucial feature of these decay rates is that they contain terms independent of the wino-wino and zino-zino mixing parameters that dominate the total rate for large gaugino masses. The decay into $W^*\tau$ is always more important than that into $Z^*\nu_\tau$ and they typically give a significant correction to the gravitino decay width, particularly for large gaugino masses. Figure \ref{3bbranchings} shows the gravitino branching ratio into the three-body final states, BR($\gravitino\to W^*\tau$)+BR($\gravitino\to Z^*\nu_\tau$), as a function of the gravitino mass for different values of $M_1$. As clearly seen in the figure,  the three-body final states give an important contribution to the decays of the gravitino. They even dominate  gravitino decays over a wide range of gaugino and gravitino masses. 

\EPSFIGURE{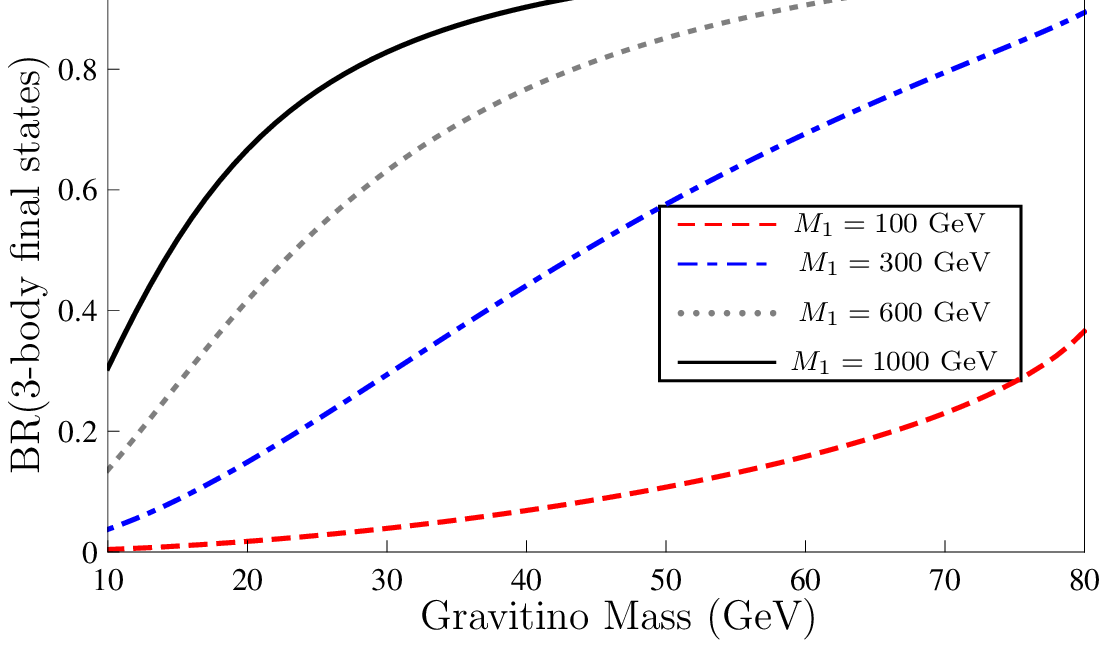}{The gravitino branching ratio into three-body final states (BR($\gravitino\to W^*\tau$)+BR($\gravitino\to Z^*\nu_\tau$)) as a function of the gravitino mass for different values of $M_1$. These branching ratios are independent of $\xi_\tau$. \label{3bbranchings}}

The total decay width of a gravitino with $\mgravitino <M_W$ will, then, be given by
\begin{equation}
\Gamma_{\text{tot}}(\gravitino)= 2(\Gamma(\gravitino\to \gamma\nu_\tau)+\Gamma(\gravitino\to \tau^+ W^{-*})+ \Gamma(\gravitino\to \nu_\tau Z^*)),
\label{eq:gwidth}
\end{equation}
where the factor two takes into account the charge conjugated final states. This decay width is  proportional to the R-parity breaking parameter $\xi_\tau^2$.  The gravitino lifetime, $\tau_{\gravitino}$, is simply the inverse of the decay width, $1/\Gamma_{\text{tot}}(\gravitino)$.

Because the indirect detection signatures of gravitino dark matter depend on the gravitino lifetime and its branching ratios, and they are both strongly affected by the three-body final states $W^*\tau$ and $Z^*\nu_\tau$, it is necessary to revisit the prospects for indirect detection of gravitinos in view of these new decay modes. At a qualitative level, it is easy to understand the novel features induced by the decays into $W^*\tau$ and $Z^*\nu_\tau$. Regarding gamma rays, the implications are twofold. On the one hand, one expects a suppression in the gamma ray line  from the two-body decay, due to the smaller branching into $\gamma\nu_\tau$ associated to a given gravitino lifetime. On the other hand, there will also be a new continuum contribution entirely due to the  three-body final states. Regarding antimatter signals, the three-body gravitino decays give rise to a non-zero positron and antiproton flux for $\mgravitino<M_W$, opening up the possibility of constraining or discovering gravitino dark matter via antimatter searches. In the next sections we will study quantitatively the implications of the three-body final states for the indirect detection of gravitino dark matter.

\section{Gamma rays from gravitino decay}
\label{sec:gammas}
Since the gravitino typically decays with sizable branching ratios into both $\gamma\nu_\tau$ and $W^*\tau+Z^*\nu_\tau$, the gamma ray spectrum from gravitino decays features a monochromatic contribution from the two-body final state and a continuum contribution from the three-body final states. The relative weight between them is determined by the decay branching ratios, which depend on $\mgravitino$ and $M_1$, whereas the overall normalization of the spectrum is controlled by the R-parity breaking parameter $\xi_\tau$.  In this section we study in detail the gamma ray spectrum from gravitino decay and we determine the constraints that can be imposed on scenarios with gravitino dark matter from recent Fermi data.

\subsection{The gamma ray spectrum}
The contribution of the two-body final state to the photon spectrum, which is simply a delta function at $E=\mgravitino/2$, has already been considered in previous works -see e.g. \cite{Bertone:2007aw,Ibarra:2007wg}. The contribution of the three-body final states $W^*\tau$ and $Z^*\nu_\tau$, on the other hand, is computed here for the first time. The photons in this case are produced after the  decay and fragmentation of the final states and come mainly from $\tau$ bremsstrahlung and neutral pion decays. To obtain the spectra we have used the event generator {\tt PYTHIA} \cite{Sjostrand:2006za}. The gamma ray spectrum from $\gravitino$ decay, $dN/dE$,  features a monochromatic and a continuum contribution and can be expressed as
\begin{equation}
\frac{dN}{dE_\gamma}=\underbrace{BR(\gravitino\to \gamma\nu)\frac{dN^{\gamma\nu}}{dE_{\gamma~}}}_\textrm{monochromatic}+ \underbrace{BR(\gravitino\to W^*\tau)\frac{dN^{W^*\tau}}{dE_{\gamma~~~}}+BR(\gravitino\to Z^*\nu_\tau)\frac{dN^{Z^*\nu}}{dE_{\gamma~~~}}}_\textrm{continuum}.
\end{equation}
Here, it is not necessary to include the inverse compton contribution to the gamma ray flux, as it is known to be negligible \cite{Buchmuller:2009xv} for the range of gravitino masses we are considering.

\DOUBLEFIGURE{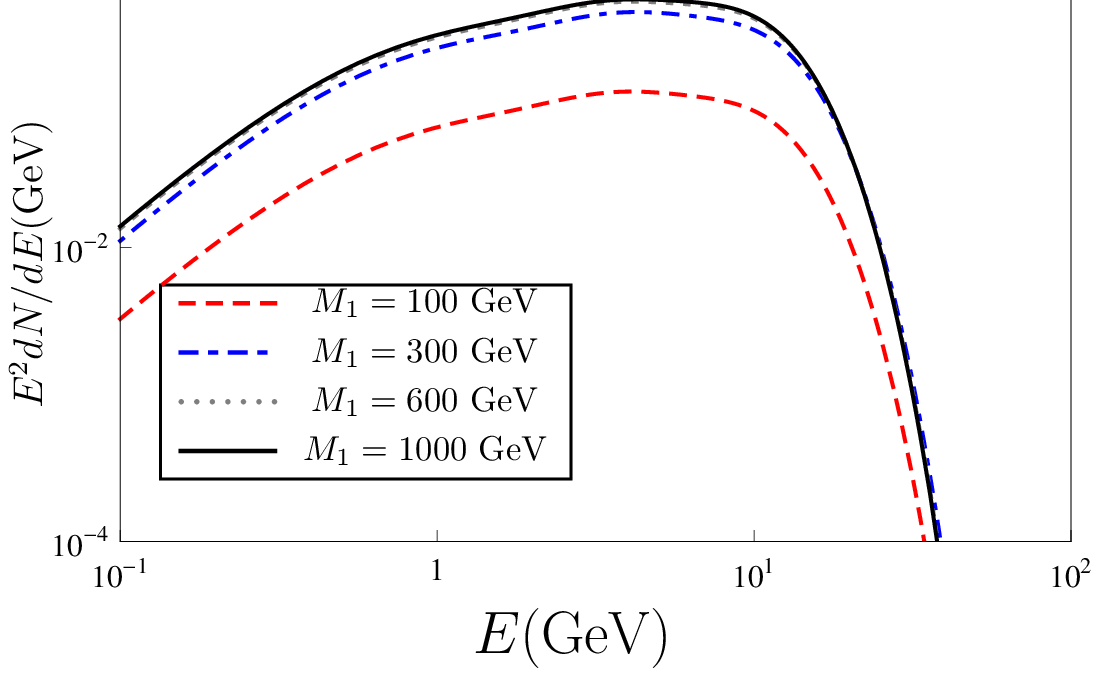,scale=0.65}{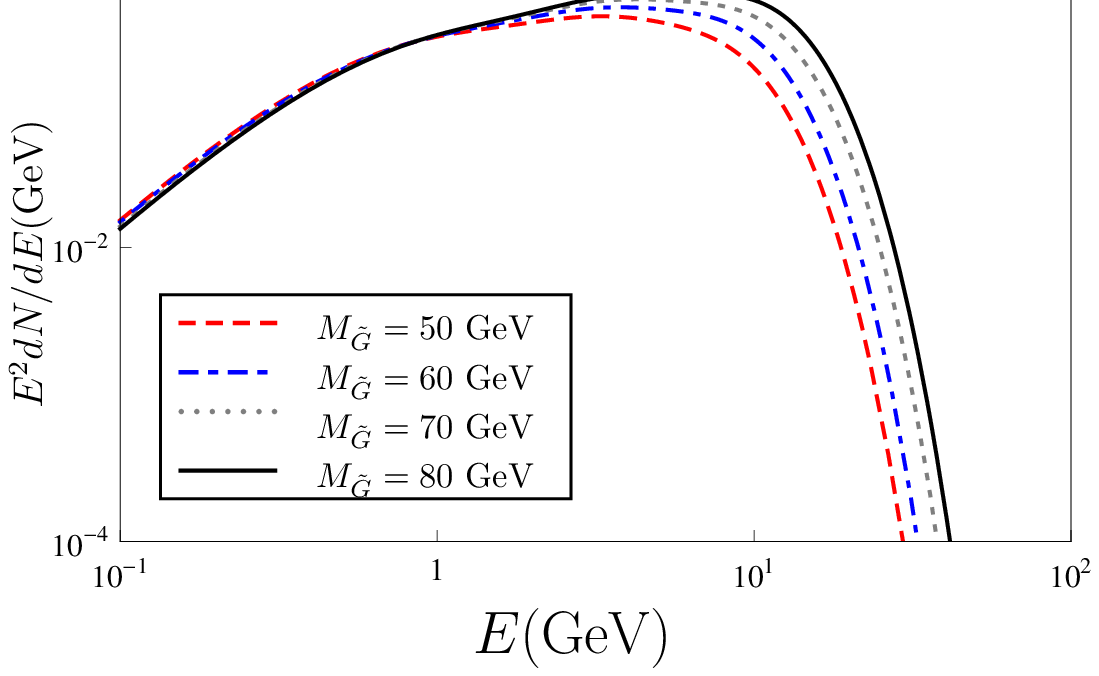,scale=0.65}{The continuum gamma ray spectrum from gravitino decay for $\mgravitino=70\gev$ and different values of $M_1$.\label{fig:gammaspecmg70}}{The continuum gamma ray spectrum from gravitino decay for $M_1=1\tev$ and different values of the gravitino mass.\label{fig:gammaspecm11000}}

To begin with, let us study the continuum contribution to the gamma ray spectrum from gravitino decay. Its dependence on gaugino and gravitino masses is illustrated in figures \ref{fig:gammaspecmg70} and \ref{fig:gammaspecm11000}. Figure \ref{fig:gammaspecmg70} shows this contribution  as a function of the energy  for $\mgravitino=70\gev$ and different values of $M_1$. Because of the higher branching ratio into $W^*\tau$ and $Z^*\nu_\tau$, the continuum contribution increases with  gaugino masses. This increase is quite important for low gaugino masses (notice the difference between the $M_1=100\gev$ and the $M_1=300\gev$ lines) but becomes irrelevant above $M_1\sim 1\tev$.  The reason is that  the branching into the three-body final states is already very close to $1$ in that region, so the spectrum hardly changes for even higher values of $M_1$.

Figure \ref{fig:gammaspecm11000} illustrates the variation of the continuum contribution  with the gravitino mass. In this figure $M_1$ was set to $1\tev$, which results, for the gravitino masses we consider, in a three-body branching larger than $90\%$ --see figure \ref{3bbranchings}.  We observe that the continuum contribution is  sensitive to the gravitino mass at high energies and that it is almost independent of it for $E\lesssim 1~\gev$.

Now that we have obtained the gamma ray spectra from gravitino decays, the next step is the computation of the expected flux at earth.  Two different populations of gravitinos contribute to such flux: gravitinos from the Milky Way and gravitinos at cosmological distances. It turns out that the dominant contribution comes from gravitinos in the Galactic halo \cite{Bertone:2007aw,Buchmuller:2009xv} so we will focus on those.  The gamma ray flux at earth from gravitinos decaying in the Milky Way halo can be computed as
\begin{equation}
\frac{d\phi_\gamma(\psi)}{dE}=\frac{2}{\mgravitino}\frac{dN}{dE_\gamma}\frac{1}{8\pi\tau_{\gravitino}}\int_{\textrm{l.o.s.}}\rho(\vec l)d\vec l\,,
\label{eq:gflux}
\end{equation}
where  $\rho(\vec l)$ is the dark matter distribution in the halo, and the integral extends over the line of sight. Notice from this equation  that the gamma ray flux from gravitino decay depends, on the particle physics side, only on the spectra, the gravitino mass and the gravitino lifetime, which is in turn determined by  $\mgravitino$, $\xi_\tau$ and $M_1$. For $\rho$ we  use the Navarro, Frenk and White (NFW) profile \cite{Navarro:1996gj},
\begin{equation}
\rho(r)=\frac{\rho_c}{r/r_c (1+r/r_c)^2},
\end{equation}
with $r_c=20~\mathrm{kpc}$ and $\rho_c=0.34~\gev~\text{cm}^{-3}$, which correspond to a local dark matter density $\rho_0=\rho(r=8.5~\mathrm{kpc})$ of $0.39~\gev~\text{cm}^{-3}$ \cite{Catena:2009mf}. To facilitate the comparison with current observations, in our numerical analysis we will be concerned only with  the high latitude region, $|b|>10^\circ$. In contrast to the Galactic Center, this region has the benefit of not being that sensitive to the assumed halo profile. For that region we find that 

\begin{equation}
\left.\frac{d\phi_\gamma}{dE}\right|_{|b|>10^\circ}=\frac{2}{\mgravitino}\frac{dN}{dE_\gamma}\frac{1}{8\pi\tau_{\gravitino}}\times 2.1\times 10^{22}~\text{GeV}/\text{cm}^2\text{str}.
\label{eq:gflux2}
\end{equation}

\DOUBLEFIGURE{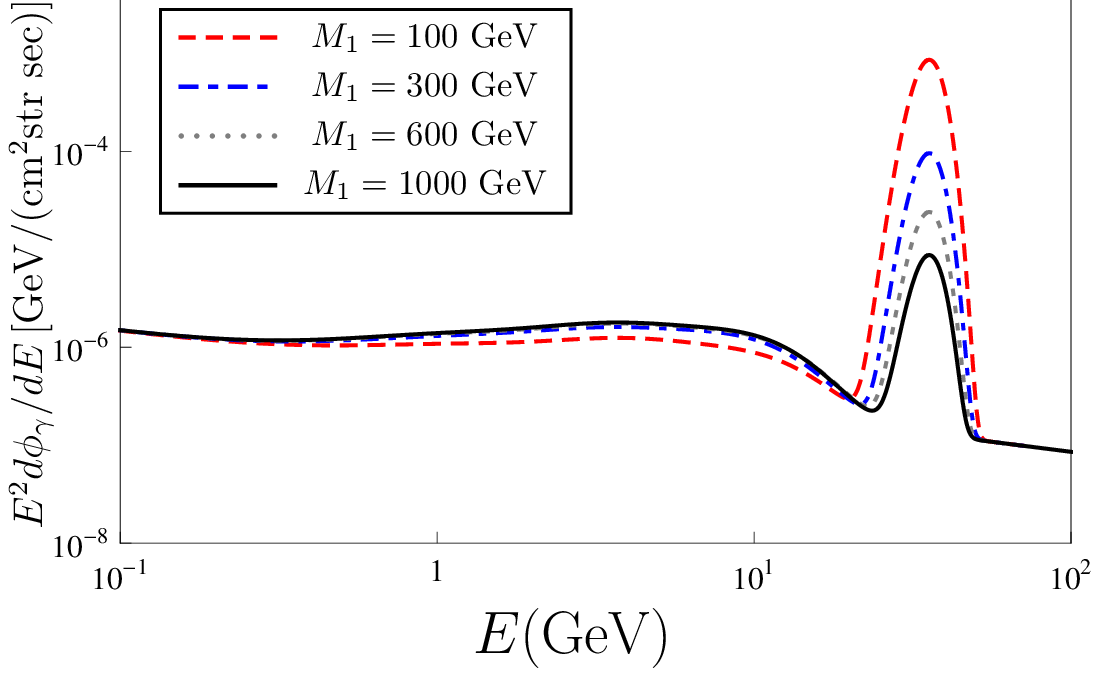,scale=0.65}{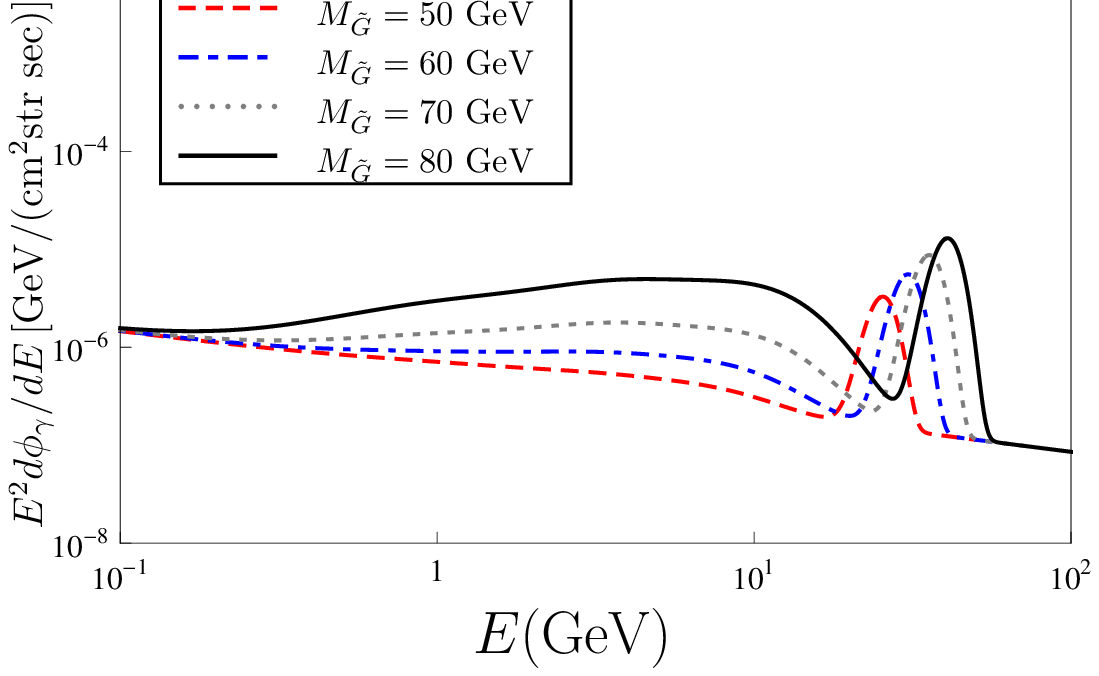,scale=0.65}{The gamma ray flux from gravitino decays as a function of the energy for $\xi_\tau=10^{-7}$, $\mgravitino=70\gev$, and different values of $M_1$.\label{gfluxmg70}}{The gamma ray flux from gravitino decays as a function of the energy for $\xi_\tau=10^{-7}$, $M_1=1\tev$, and different values of $\mgravitino$.\label{gfluxm11tev}}

Figures \ref{gfluxmg70} and \ref{gfluxm11tev} show the dependence of the gamma ray flux from gravitino decay on $\mgravitino$ and $M_1$. The monochromatic contribution has been convoluted with a $10\%$ gaussian energy resolution, comparable to that expected from FERMI.  From these two figures we can see that the monochromatic and continuum contributions are clearly discernible. The former dominates the flux at high energies, $E\sim \mgravitino/2$, whereas the latter does it at somewhat lower energies. Figure \ref{gfluxmg70} shows that the height of the monochromatic contribution is quite sensitive to the value of $M_1$. The continuum contribution, on the other hand, is only slightly affected by $M_1$. As expected from the gravitino branching ratios, the larger the gaugino masses the smaller the line signal and the larger the continuum. From figure \ref{gfluxm11tev} we see that $\mgravitino$ not only determines the position of the line signal but also the height of the continuum contribution, being larger for heavier gravitinos. If such a flux were observed, $\mgravitino$ could be extracted from the position of the peak, while the gaugino mass $M_1$ could be determined from the  relative weight between the monochromatic and continuum contributions. 

In previous studies on the indirect detection of gravitino dark matter only the two-body decays of the gravitino were considered. That is, the branching ratio into $\gamma\nu_\tau$ was assumed to be equal to $1$ for $\mgravitino<M_W$ --see e.g. \cite{Bertone:2007aw,Ibarra:2007wg}.  It is useful, therefore, to compare the correct gamma ray flux expected from gravitino decays (two- and three-body decays included) with that obtained in previous works (only the two-body decay). Figure \ref{23comparison} shows the predicted gamma ray flux in both cases. For that figure we set  $\mgravitino=70\gev$, $M_1=1\tev$,  and $\tau_\gravitino=8\times10^{24}~\text{s}$. The correct flux features a much smaller monochromatic contribution and the entirely new continuum contribution. Clearly, to  reliably assess the indirect detection prospects of gravitino dark matter via gamma rays, the three-body decays of the gravitino must necessarily be taken into account. 

\EPSFIGURE{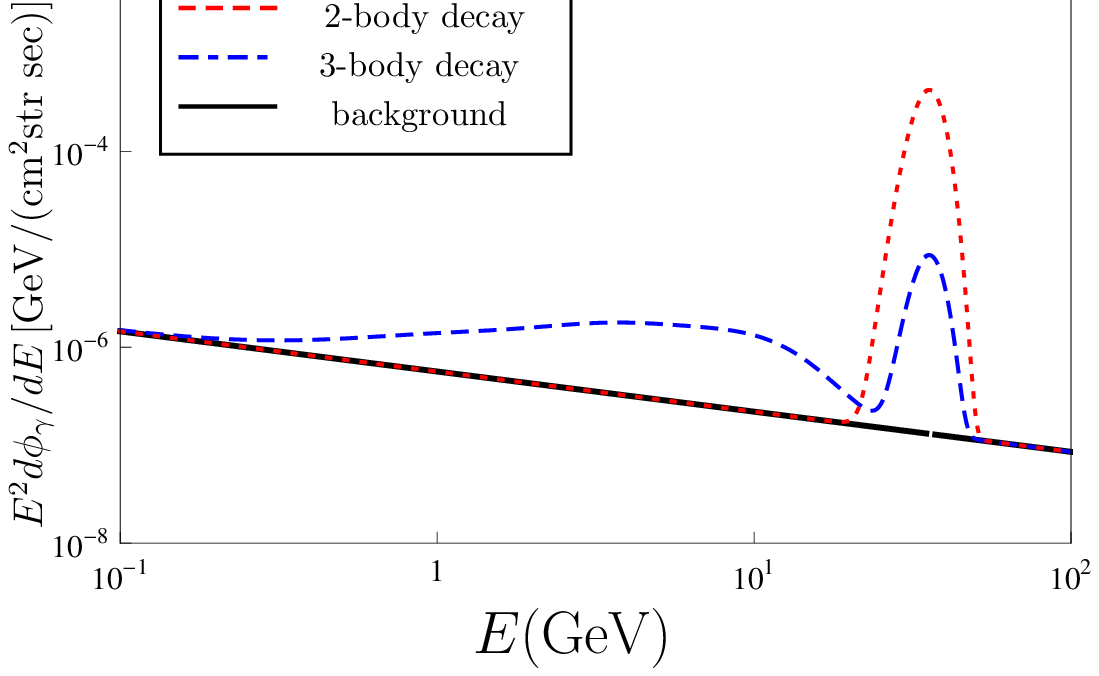}{A comparison between the correct gamma ray flux expected from  a  decaying gravitino (dashed line) and the corresponding flux obtained including  2-body final states only (dotted line).  For this figure we set $\mgravitino=70\gev$, $M_1=1\tev\gev$,  and $\tau_\gravitino=8\times10^{24}~\text{s}$. The gamma ray background is also shown (solid line). \label{23comparison}}

Next, the gamma ray fluxes we just obtained are used in conjunction with recent FERMI data to derive  new constraints on scenarios with gravitino dark matter  and  bilinear R-parity violation. 
\subsection{Constraints on gravitino dark matter from FERMI data}
Recently, the Fermi collaboration published two results that can be used to constrain the parameter space of gravitino dark matter. On the one hand, they searched for photon lines with energies between $30\gev$ and $200\gev$ and derived a constraint on the annihilation or decay rate of dark matter particles into monochromatic photons \cite{Abdo:2010nc} \footnote{A new analysis using two years of Fermi data has just been published \cite{newglines}.}. Since the two-body decay of the gravitino produces a gamma ray line at $\mgravitino/2$, we can apply such constraints to our scenario. In addition, Fermi also measured the diffuse gamma ray emission for the Galactic latitude range $|b|>10^\circ$ and energies between $100\mev$ and $100\gev$, and found it to be compatible with the theoretical expectation of the  diffuse background obtained in  models of
cosmic ray propagation in our Galaxy \cite{Abdo:2010nz}. These measurements have already been used to constrain different models of decaying or annihilating dark matter, see e.g. \cite{otherconstraints}. As we have seen, gravitino decays give an additional contribution to the gamma ray flux in that energy range, so we must make sure that the gravitino contribution is small enough to remain compatible with data. Here, we will use those two results to derive constraints on the value of $\xi_\tau$ and on the gravitino lifetime for different values of $M_1$ and $\mgravitino$.

\EPSFIGURE{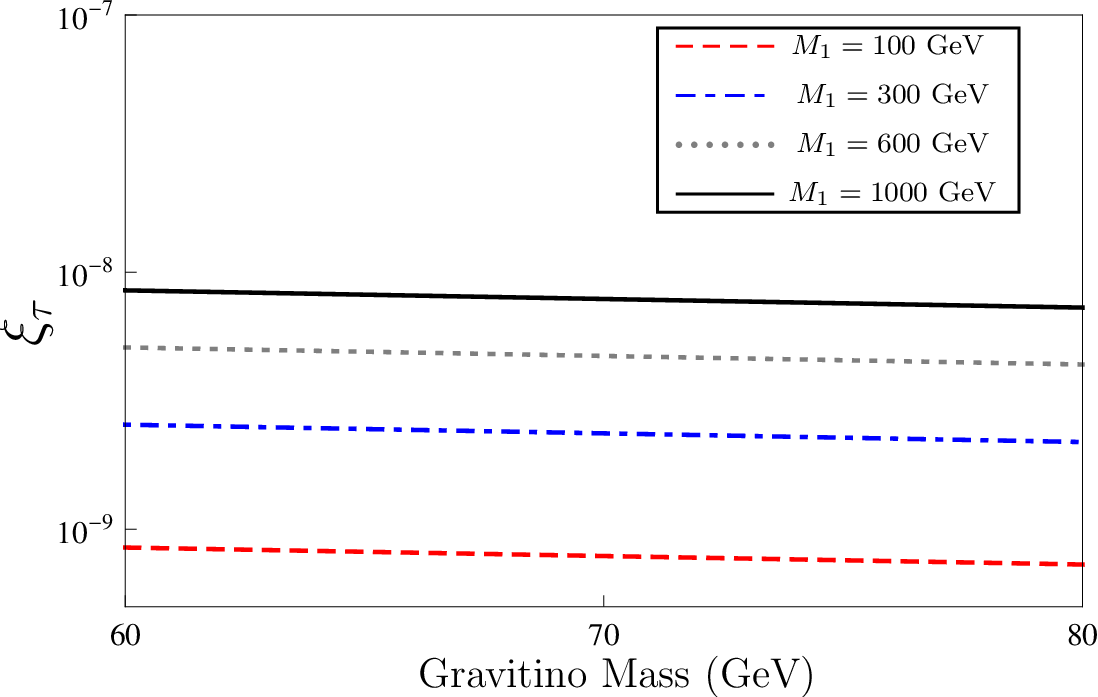,scale=1.0}{The constraint on $\xi_\tau$ derived from the gamma ray line searches by Fermi. The region above the lines is already ruled out by Fermi data for the respective value of $M_1$. Notice that the FERMI gamma ray line searches \cite{Abdo:2010nc} do not constrain gravitino masses below $60 \gev$. \label{xitaumgline}}

In figure \ref{xitaumgline} we show the exclusion regions, in the plane ($\mgravitino,\xi_\tau$), derived from the  Fermi bounds on a gamma ray line from dark matter decay. Notice that, in our scenario, those bounds  apply only to the range $60\gev<\mgravitino<80\gev$. To obtain the exclusion regions we simply rescaled the bound from Table I in \cite{Abdo:2010nc} to take into account that only one photon is emitted in a gravitino decay. As the figure illustrates, the line constraint does not depend much on the gravitino mass --at least in the range of interest to our discussion. It does depend on gaugino masses. It is one order of magnitude weaker for $M_1=1\tev$ than for $M_1=100\gev$.    

\EPSFIGURE{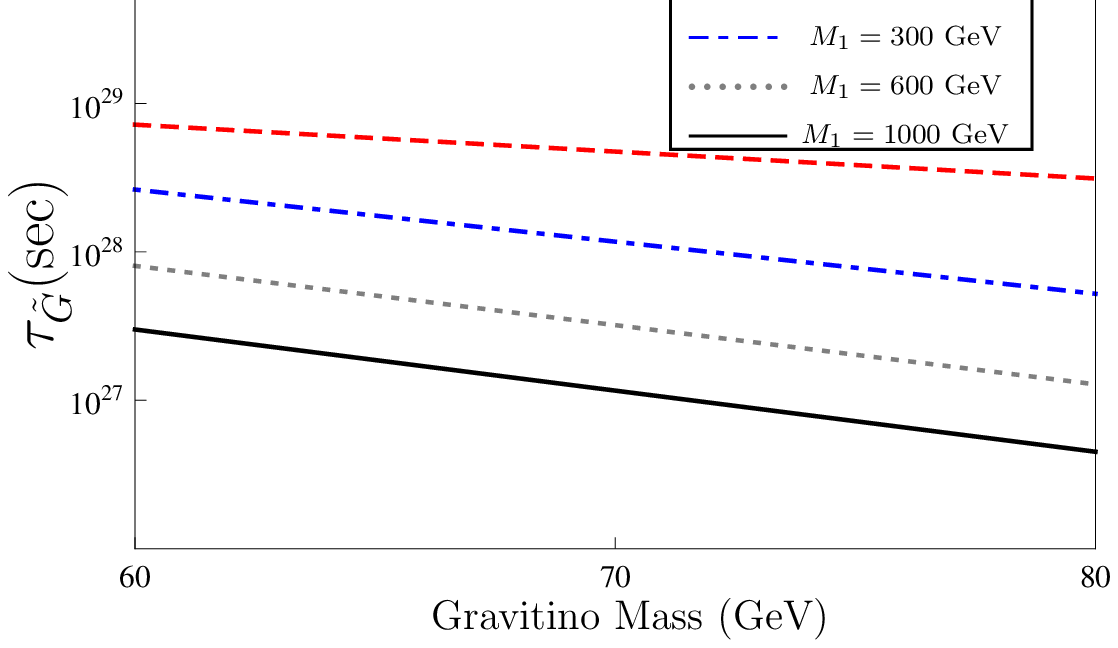,scale=1.0}{The constraint on the gravitino lifetime derived from the gamma ray line searches by Fermi. The region below the lines is already ruled out by Fermi data for the respective value of $M_1$. Notice that the FERMI gamma ray line searches \cite{Abdo:2010nc} do not constrain gravitino masses below $60 \gev$. \label{gltline}}

We can also write these results as constraints on the gravitino lifetime rather than on $\xi_\tau$. Figure \ref{gltline} shows exclusion regions in the plane ($\mgravitino, \tau_{\gravitino}$) for different values of $M_1$. The area below the lines is excluded by Fermi data for the respective value of $M_1$. We see that the constraint on the gravitino lifetime lies approximately between $\tau_\gravitino>10^{29}$ s for $M_1=100$ and $\tau_\gravitino>10^{27}$ s for $M_1=1~\tev$, with a slight dependence on the gravitino mass. As before, this bound on $\tau_\gravitino$ from Fermi line searches applies only to gravitino masses between $60$ and $80~\gev$.

A bound valid over a wider mass range can be obtained from the isotropic diffuse Fermi data. The spectrum of the extragalactic background measured by Fermi is compatible with a power law with index $\gamma=2.41\pm0.05$ \cite{Abdo:2010nz}. To obtain the exclusion constraints from the isotropic diffuse gamma-ray data \cite{Abdo:2010nz}, we compare 
 the expected signal plus background with Fermi data for each energy bin. A model is considered to be excluded if the signal plus background deviates  from the data  by more than 3$\sigma$ in any energy bin. Exclusion zones in the plane $(m_{\tilde{G}},\xi_\tau)$ are obtained by increasing $\xi_\tau$ until that condition is satisfied. 

\EPSFIGURE{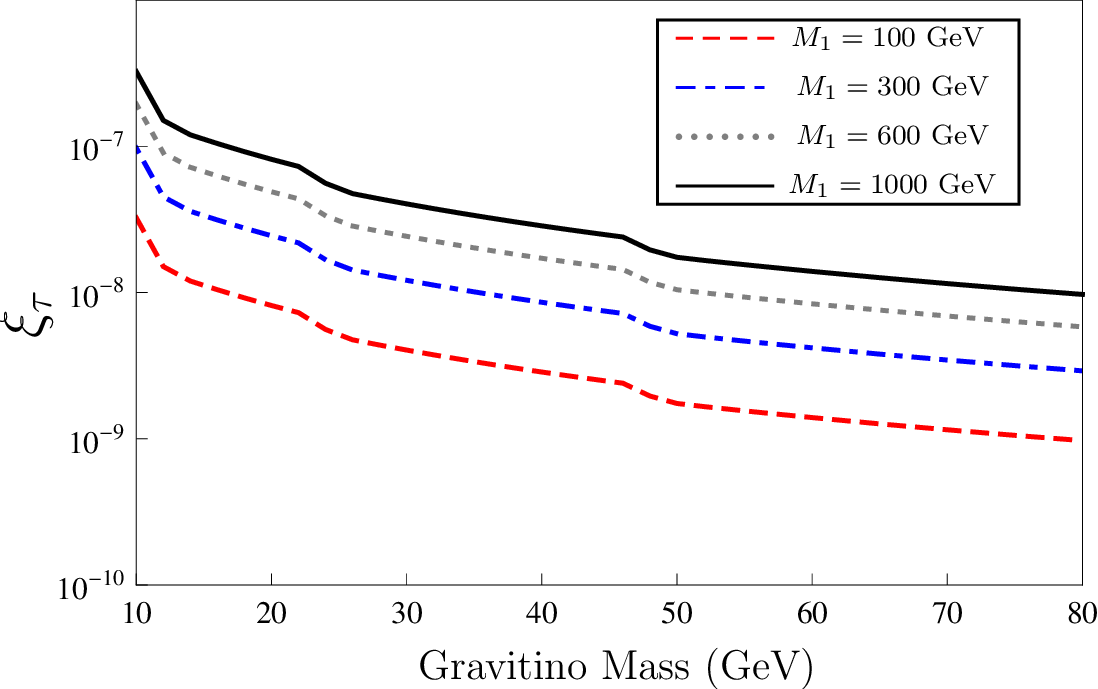,scale=1.0}{The maximum allowed values of $\xi_\tau$ as a function of the gravitino mass for different values of  $M_1$. The region above the lines is already ruled out by Fermi data.\label{xitaumg}}

Figure \ref{xitaumg} shows the constraints on $\xi_\tau$ derived from the diffuse Fermi data. It displays exclusion lines in the plane ($\mgravitino,\xi_\tau$) for different values of $M_1$. The region above the lines is excluded for the respective value of $M_1$. The step-like features observed in the figure are the result of   the line feature in the spectrum moving through the different FERMI energy bins.
Notice that the constraint on $\xi_\tau$ is more stringent at smaller gravitino masses and that it strongly depends on gaugino masses. It is, in fact, about one order of magnitude stronger for  $M_1=100~\gev$ than for $M_1=1~\tev$. This dependence on $M_1$ clearly indicates that the monochromatic contribution  is dominantly responsible for the constraint, in agreement with the fluxes we derived previously --see figures \label{gfluxmg70,gfluxm11tev}. Comparing figures \ref{xitaumg} and \ref{xitaumgline} we notice that, for gravitino masses between $60$ and $80~\gev$,  the strongest constraint on $\xi_\tau$ comes from the line searches rather than the diffuse emission data. At smaller gravitino masses, where there are no bounds from line searches, the allowed values of $\xi_\tau$ are determined by the diffuse Fermi data.

\EPSFIGURE{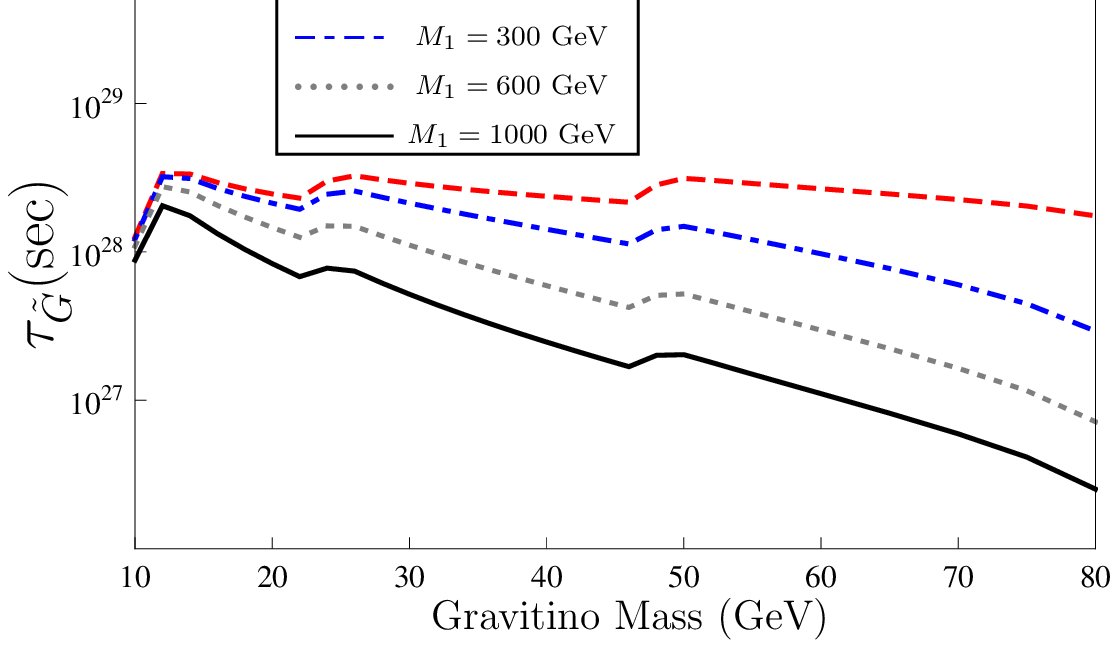,scale=1.0}{The minimum allowed values of $\tau_{\gravitino}$ as a function of the gravitino mass for different values of  $M_1$. The region below the lines is already ruled out by Fermi data.\label{taumg}}

The bound on $\xi_\tau$ we have just obtained can be translated, via equation (\ref{eq:gwidth}), into a constraint on the gravitino lifetime, as illustrated by Figure \ref{taumg}. In it, the region below the lines is excluded by the diffuse Fermi data.  Let us first emphasize that the dependence of these constraints on gaugino masses is entirely due to the three-body final states $W^*\tau$ and $Z^*\nu_\tau$. If only two-body final states were considered there would have been a unique line in this figure.  From the figure we see that for $M_1=100~\gev$ the gravitino lifetime should be larger than about $2\times 10^{28}$ s, with only a mild dependence on the gravitino mass. For $M_1=1~\tev$, the constraint can be up to two orders of magnitude weaker, depending on the  gravitino mass. As expected, at low gravitino masses  the effect of the three-body final states is smaller (see figure \ref{3bbranchings}), giving rise to a weaker dependence of the constraint with gaugino masses.

To summarize, we have seen that the three-body decays of the gravitino considerably alter the  gamma ray spectrum expected from  $\gravitino$ decays and the constraints that can be derived from gamma ray data.  
\section{Antimatter searches}
\label{sec:anti}
Antimatter searches are another promising way to indirectly detect dark matter. The positrons and antiprotons produced in gravitino decays originate entirely in the $W^*\tau$ and $Z^*\nu_\tau$ final states and had not, in consequence, been considered before in the literature. In this section, we obtain the positron and antiproton spectra from gravitino decay and compute their expected fluxes at earth.

\subsection{Positrons}
The positrons from gravitino decay into $W^*\tau$ and $Z^*\nu_\tau$ originate in the decay of the $\tau^\pm$ and in the fragmentation of the virtual gauge bosons, mainly via $\pi^+$ decay. They  give rise to a continuum spectrum of positrons, which we  obtain using the event generator {\tt PYTHIA} \cite{Sjostrand:2006za}. Figure \ref{epspectrum} shows the positron spectrum (before propagation) resulting from gravitino decay for $\mgravitino=70~\gev$ and different values of $M_1$. At high energies, $E^2{dN_{e^+}}/{dE}$ is observed to increase with energy, indicating that ${dN_{e^+}}/{dE}$ is a slowly decreasing function in that region --a result that can be attributed to the positrons from tau decay. As the gauginos become heavier,  the branching  ratio into three-body final states gets larger and so does the positron flux. In fact, it is observed to be about four times larger for $M_1=1~\tev$ than for $M_1=100~\gev$ over the whole energy range. For $M_1\sim 1~\tev$, the branching into $W^*\tau+Z^*\nu_\tau$ is almost $1$ and the positron flux cannot longer increase with gaugino masses.    
\begin{center}
\EPSFIGURE{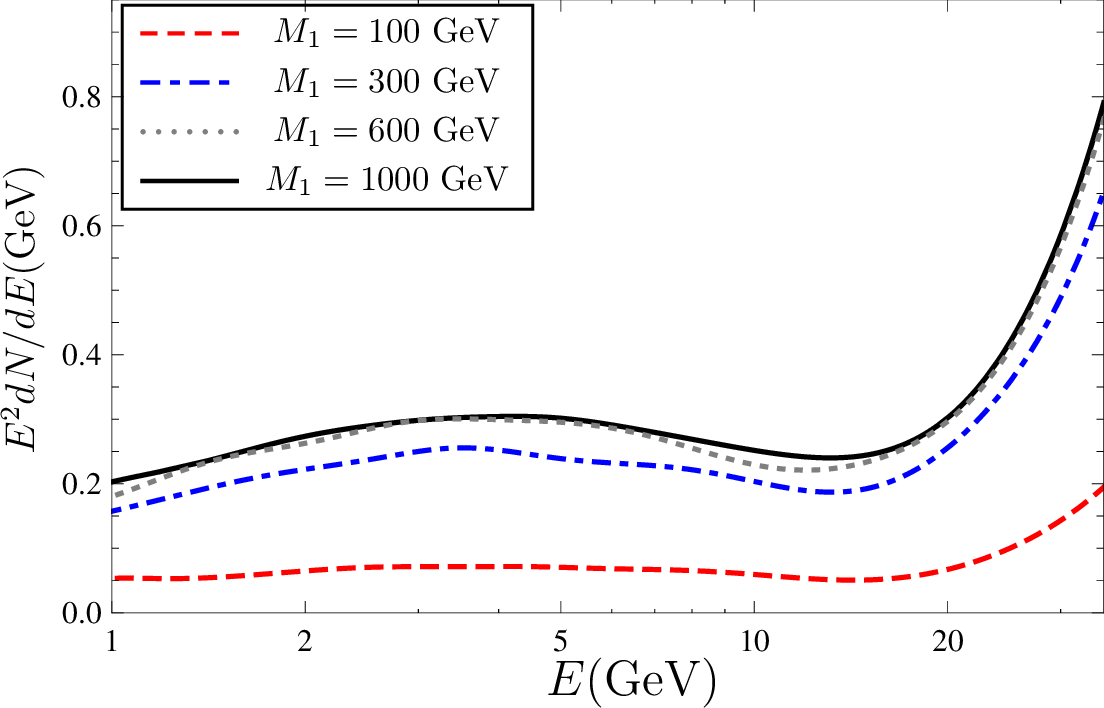,scale=1.0}{The positron spectrum (before propagation) from gravitino decay for different values of $M_1$. For this figure we set $\mgravitino=70~\gev$.\label{epspectrum}}
\end{center}

Unlike gamma rays, which travel directly to us, positrons are affected by propagation effects in the Galaxy.   A careful discussion of these effects and of the theoretical uncertainties involved was presented in \cite{Delahaye:2007fr}. The important point for us is that propagation effects can be taken into account by solving, for the positron number density $f=dN_{e^+}/dE$,  the following  stationary diffusion-loss equation:
\begin{equation}
-K(E)\nabla^2f-\frac{\partial}{\partial E}(b(E) f)=Q_{e^+}\,,
\label{eq:positrons}
\end{equation}
where $K(E)=K_0(E/\gev)^\delta$ is the diffusion coefficient and $b(E)=E^2/(\gev\cdot \tau_E)$ is the energy loss coefficient, with $\tau_E=10^{16}$ s. They describe, respectively, transport through the turbulent magnetic field and energy losses due to synchrotron radiation and inverse compton scattering on CMB photons and on galactic starlight. For positrons originating in gravitino decays, the source term, $Q_{e^+}$,  is given by 
\begin{equation}
Q_{e^+}(r,E)=\frac{\rho(r)}{\mgravitino}\sum_i\Gamma_i\frac{dN_{e^+}^i}{dE}\,,
\end{equation}
where the sum is over the different decay modes that can generate positrons. That is, over $W^*\tau$ and $Z^*\nu_\tau$. 

Equation (\ref{eq:positrons}) is solved within a diffusive region with the shape of a flat cylinder. Its radius $R$ is $20~\mathrm{kpc}$ and its height is $2L$ in the $z$ direction. On the surface of this cylinder, the positron density $f$ vanishes. The solar system is  located at $(r,z)=(8.5~\mathrm{kpc},0)$. Thanks to these simplifying assumptions, the positron flux at earth can be written in a semi-analytical form (see e.g. \cite{Ibarra:2008qg,Cirelli:2008id}. In this framework, $f$ will depend on the propagation parameters $\delta$, $K_0$ and $L$, which are constrained by cosmic ray data. Here, we will simply adopt the so-called MED propagation model \cite{Delahaye:2007fr} ($\delta=0.70\,, K_0=0.0112~\mathrm{kpc^2/Myr}\,,L=4~\mathrm{kpc}$), which provides the best fit to the Boron-to-Carbon (B/C) ratio.

To compare with experimental data, it is better to use, rather than the positron flux, the positron fraction: the ratio between the  positron flux and the electron+positron flux. To compute the positron fraction, we need to know not only the positron flux from gravitino decays but also the background electron and positron fluxes.  The interstellar background fluxes of electrons and positrons can be parameterized as \cite{Ibarra:2009dr}
\begin{align}
\Phi_{e^-}^{\text{bkg}}(E)&=\left(\frac{82.0 \epsilon^{-0.28}}{1+0.224\epsilon^{2.93}}\right) \mathrm{GeV^{-1}m^{-2}s^{-1}sr^{-1}}\,,\label{eq:bkg1}\\
\Phi_{e^+}^{\text{bkg}}(E)&=\left(\frac{38.4 \epsilon^{-4.78}}{1+0.0002\epsilon^{5.63}}+24.0\epsilon^{-3.41}\right) \mathrm{GeV^{-1}m^{-2}s^{-1}sr^{-1}}\,.
\label{eq:bkg2}
\end{align}
with $\epsilon=E/\gev$. These expressions reproduce to an excellent approximation the results obtained for a GALPROP conventional model -- Model 0 in \cite{Grasso:2009ma}-- that is compatible with the gamma-ray diffuse emission spectrum measured by Fermi at intermediate Galactic latitudes. Finally, it must be taken into account that for energies smaller than about $10~\gev$ the electron and positron fluxes at the top of the atmosphere (the measured ones) can differ, because of solar modulation effects, from the interstellar fluxes. In the force field approximation \cite{Gleeson:1968zz}, the fluxes at the top of the atmosphere are related to the intersterllar fluxes by the following equation:
\begin{equation}
\Phi_{e^\pm}^{\mathrm{TOA}}(E_{\mathrm{TOA}})=\frac{E_{\mathrm{TOA}}^2}{E_{\mathrm{IS}}^2}\Phi_{e^\pm}^{\mathrm{IS}}(E_{\mathrm{IS}})
\end{equation}
where $E_\mathrm{IS}$, $E_\mathrm{TOA}$ are the electron/positron fluxes at the heliospheric boundary and at the top of the Earth's atmosphere respectively. They are related with each other via the solar modulation parameter, $\phi_F$, as $E_\mathrm{IS}=E_\mathrm{TOA}+\phi_F$. In our calculations we use $\phi_F=500~$MV, which is characteristic of the minimum of solar cyclic activity. We now have all the ingredients required to compute the positron fraction in our scenario and to compare it with present data. 

%\begin{center}
\EPSFIGURE{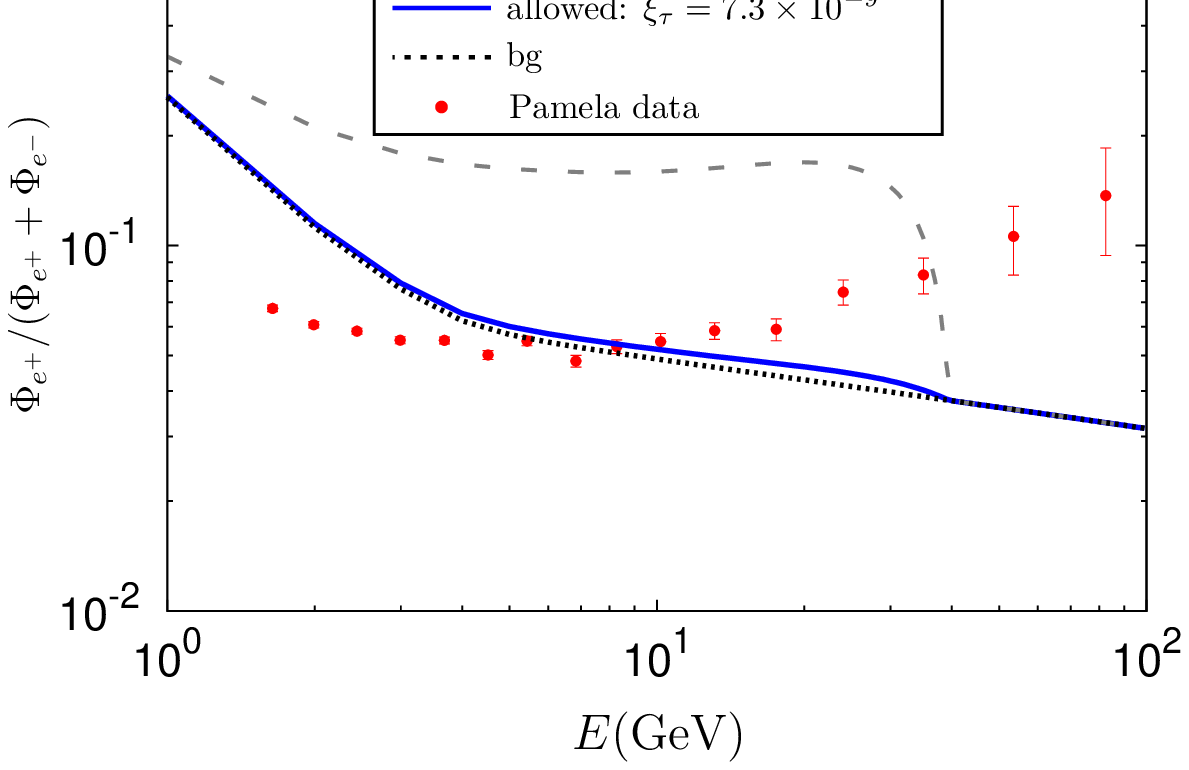,scale=1.0}{The positron fraction as a function of the energy for $\mgravitino=80~\gev$ and $M_1=1\tev$. The interstellar background and the PAMELA data are also shown. Two different  values of $\xi_\tau$ are illustrated. For $\xi_\tau=5\times 10^{-8}$ (dashed gray line) the positron fraction is clearly distinguishable from the background but  such value of $\xi_\tau$ is not compatible with gamma ray data. The solid line shows the positron fraction computed for the largest value of $\xi_\tau$ allowed by the gamma ray constraints ($7.3\times 10^{-9}$). In that case,  the positron signal from gravitino decay is indistinguishable from the background.\label{efraction}}
%\end{center}
Figure \ref{efraction} shows the predicted positron fraction as a function of the energy for $\mgravitino=80~\gev$, $M_1=1~\tev$, and two different values of $\xi_\tau$. In addition, the PAMELA measurements \cite{Adriani:2008zr} and the predicted background, from equations (\ref{eq:bkg1}) and (\ref{eq:bkg2}), are also shown. As is well known, the PAMELA data indicates an excess of positrons over the expected background \cite{Adriani:2008zr}. The inclusion of the three-body final states $W^*\tau$ and $Z^*\nu_\tau$ does not help in resolving that discrepancy, as they affect the positron flux only at relatively low energies, $E<30~\gev$. From the figure we see that for $\xi_\tau=5\times 10^{-8}$, gravitino decays  would produce a significant deviation in the positron fraction. Unfortunately, such large values of $\xi_\tau$ are not compatible with gamma ray constraints --see figure \ref{xitaumg}. The largest value  allowed by such data is about one order of magnitude smaller, $\xi_\tau=7.3\times 10^{-9}$.  As illustrated in figure \ref{efraction}, such a small $\xi_\tau$ does not give rise to an observable signal. A result that, as we have verified, is  independent of the gaugino or gravitino masses used. In conclusion, the range of parameters that is compatible with gamma ray observations yields a negligible flux of positrons from gravitino decays.  

\subsection{Antiprotons}
Gravitino decays into $W^*\tau$ and $Z^*\nu_\tau$ produce antiprotons  via gauge bosons fragmentation.  We have used {\tt PYTHIA} to obtain the predicted antiproton spectrum for several values of gaugino and  gravitino masses. Figure \ref{pbarspec} shows the antiproton spectrum (before propagation) for $\mgravitino=70~\gev$ and different values of $M_1$. The number of antiprotons produced at a given energy  is observed to increase with gaugino masses thanks to the larger branching ratio into the three-body final states. At $10\gev$, for instance, $dN_{\bar p}/dE$ is about four times larger for $M_1=1\tev$ than for $M_1=100\gev$. As before, this enhancement is saturated at $M_1\sim 1\tev$, above which the antiproton flux no longer increases with $M_1$. In contrast with the positron spectrum, the antiproton spectrum decreases at high energies rather quickly. 
 
\begin{center}
\EPSFIGURE{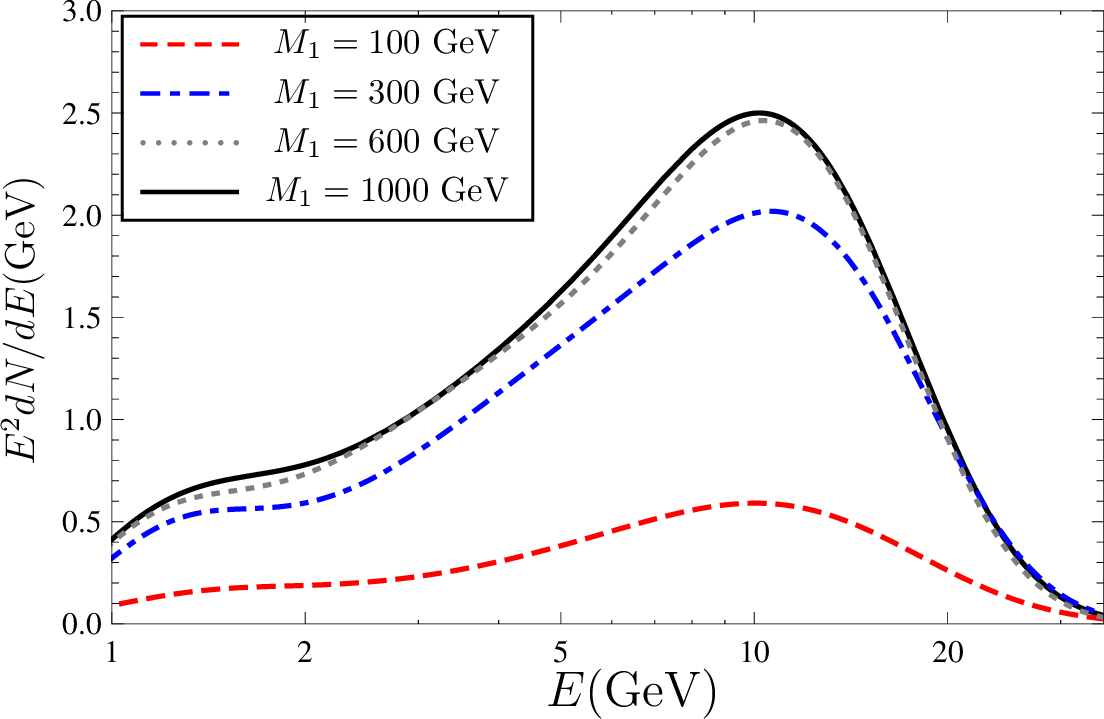,scale=1.0}{The antiproton spectrum  (before propagation) as a function of the energy for different values of $M_1$.\label{pbarspec}}
\end{center}

The propagation of antiprotons in the Galaxy is  described by a diffusion equation formally analogous to that for positrons, equation (\ref{eq:positrons}). Under certain conditions, see \cite{Cirelli:2008id} for details, the equation can be solved analytically to obtain
\begin{equation}
\Phi_{\bar p}(T,\vec r)=\frac{v_{\bar p}}{4\pi}\frac{\rho}{\mgravitino}R(T)\sum_i\Gamma_i\frac{dN_{\bar p}^i}{dT}\,,
\end{equation}
where $i$ runs over the three-body final states $W^*\tau$ and $Z^*\nu_\tau$, $T=E-m_{\bar p}$ is the antiproton kinetic energy, and $v_{\bar p}$ is the antiproton velocity. The function $R(T)$ encodes all the astrophysics and depends, in particular, on the propagation parameters.  In our calculations, we will use the MED propagation model \cite{Donato:2003xg}, for  which $\delta=0.70$, $K_0=0.0112~\mathrm{kpc^2/Myr}$, $L=4~\mathrm{kpc}$, and $V_{\text{conv}}=12~\mathrm{km/s}$. It must be kept in mind, however, that the antiproton flux is very sensitive to the choice of propagation parameters. Indeed, an order of magnitude variation with respect to the result for the MED model seems to be typical. For that reason, we will not try to derive constraints from antiproton data but simply illustrate how large the antiproton flux from gravitino decay can be.

Before doing so, we must take into consideration that solar modulation effects can modify the observed antiproton flux  at low energies. In the force field approximation \cite{Gleeson:1968zz}, the antiproton flux at the top of the Earth's atmosphere ($\Phi_{\bar p}^\mathrm{TOA}$) is related to the interstellar antiproton flux ($\Phi_{\bar p}^\mathrm{IS}$) by
\begin{equation}
\Phi_{\bar p}^\mathrm{TOA}(T_\mathrm{TOA})=\frac{2m_pT_\mathrm{TOA}+T_\mathrm{TOA}^2}{2m_pT_\mathrm{IS}+T_\mathrm{TOA}^2}\Phi_{\bar p}^\mathrm{IS}(T_\mathrm{IS})
\end{equation}
where  $T_\mathrm{IS}$, $T_\mathrm{TOA}$ are the antiproton kinetic energies respectively at the heliospheric boundary and at the top of the Earth's atmosphere. They are related as $T_\mathrm{IS}=T_\mathrm{TOA}+\phi_F$, with $\phi_F$ being the solar modulation parameter, which we take equal to $500$ MV. 

 \EPSFIGURE{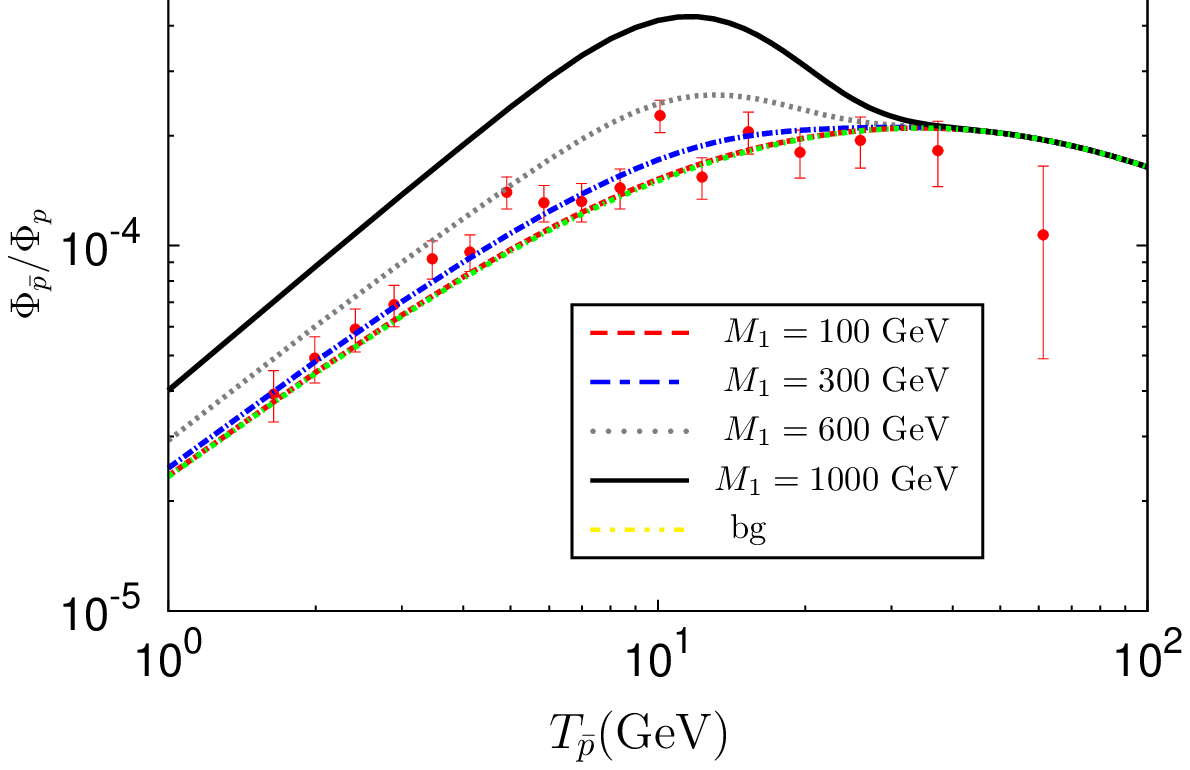,scale=1.0}{The antiproton to proton ratio as a function of the energy for different values of $M_1$. In this figure we set $\mgravitino=70~\gev$ and the value of $\xi_\tau$  is the maximum allowed by the gamma ray constraints ($\xi_\tau=7.9\times10^{-10}, 2.4\times 10^{-9}, 4.7\times 10^{-9}, 7.9\times10^{-9}$ for $M_1=100, 300, 600, 1000$ GeV, respectively).\label{pbarratio70}}
 
In order to compare with available data, we must compute the antiproton to proton ratio rather than simply the antiproton flux. To do so, we need to know the proton and antiproton backgrounds. The antiproton background, taken from \cite{Bringmann:2006im,Cirelli:2008id}, can be parametrized as
\begin{equation}
\Phi_{\bar{p}}^{\text{bkg}}=10^{-1.64+0.07\log_{10}T-\log^2_{10}T-0.02\log^3_{10}T+0.028\log^4_{10}T}.
\end{equation}
whereas the proton background is given by \cite{Bai:2009ka}
\begin{equation}
\Phi_{p}^{\text{bkg}}=e^{6.88-0.828\log T-0.382\log^2 T+0.014\log^3 T+0.012\log^4 T-0.00097\log^5 T},
\end{equation}
with $T$ in GeV and the fluxes in $\text{m}^{-2}\text{s}^{-1}\text{str}^{-1}\text{GeV}^{-1}$. The proton and antiproton fluxes are, With them and the predicted antiproton flux from gravitino decays, we can compute the expected antiproton-to-proton ratio at earth and compare it with the recent PAMELA measurements \cite{Adriani:2008zq}.

\EPSFIGURE{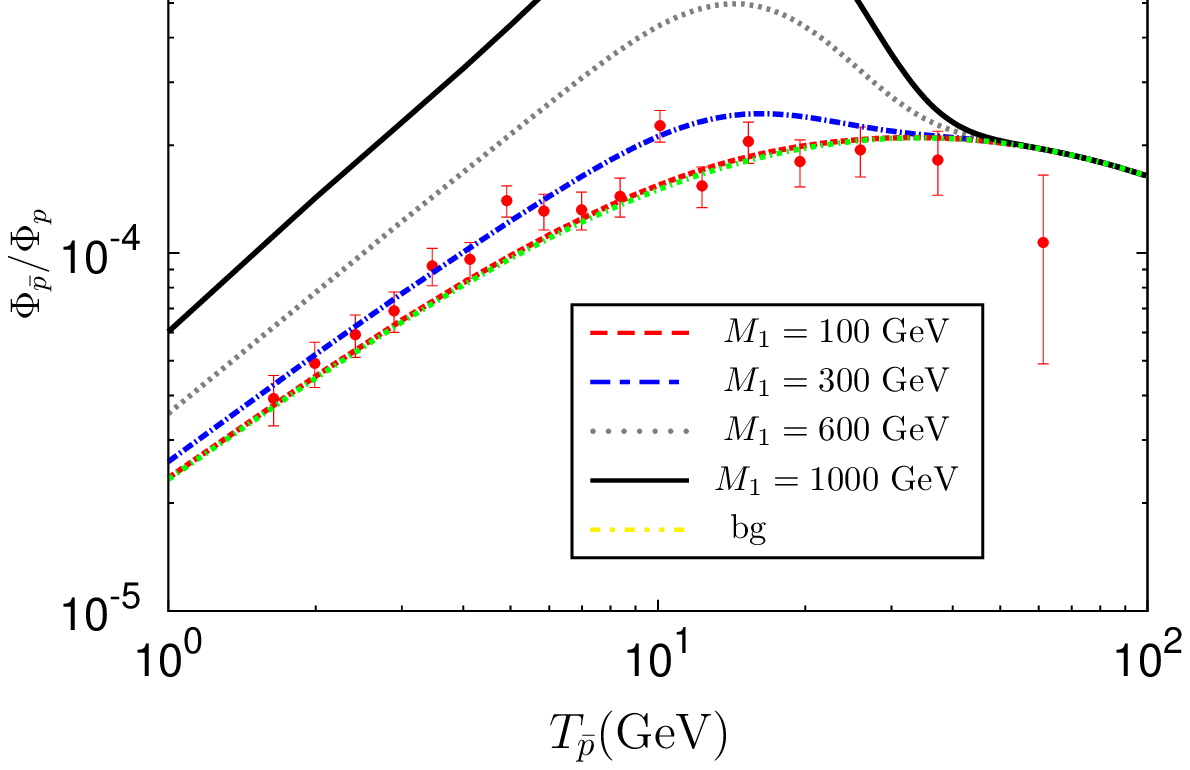,scale=1.0}{The antiproton to proton ratio as a function of the energy for different values of $M_1$. In this figure we set $\mgravitino=80~\gev$ and the value of $\xi_\tau$  is the maximum allowed by the gamma ray constraints ($\xi_\tau=7.3\times10^{-10}, 2.2\times 10^{-9}, 4.4\times 10^{-9}, 7.3\times10^{-9}$ for $M_1=100, 300, 600, 1000$ GeV, respectively).\label{pbarratio80}}

Figure \ref{pbarratio70} shows the $\bar p/p$ ratio as a function of the kinetic energy for $\mgravitino=70~\gev$ and different values of $M_1$. For comparison, the PAMELA antiproton data and the predicted background are also shown. For each $M_1$  the value of $\xi_\tau$ used is  the maximum that is  allowed by the gamma ray constraint --see figure \ref{xitaumg}. Notice that the different lines are in principle distinguishable from one another and from the background. For gravitino masses closer to $M_W$, the antiproton flux is even larger, as shown in figure \ref{pbarratio80}. It is analogous to figure \ref{pbarratio70} but for $\mgravitino=80~\gev$. Notice that, for $M_1=1~\tev$, the antiproton-to-proton ratio  deviates significantly from the  PAMELA data points in both figures. For the MED propagation model we use, such models are clearly  incompatible with present data. In the future, a better knowledge of the  propagation parameters may allow us to use antiproton measurements to constrain the supersymmetric parameters directly. As indicated by figures \ref{pbarratio70} and \ref{pbarratio80}, the antiproton signal from gravitino decay can be significant even if the constraints from gamma rays are taken into account.

\section{Conclusions}
In the context of supersymmetric models with gravitino dark matter and bilinear R-parity violation, we have studied the indirect detection prospects of gravitinos for $\mgravitino<M_W$. The main novelty in our analysis is the inclusion of  gravitino decays  into the three-body final states $W^*\tau$ and $Z^*\nu_\tau$,  which were recently shown to modify in a significant way the gravitino lifetime and its branching ratios. As a result of these new contributions, the expected fluxes of gamma rays, positrons and antiprotons  from gravitino decays are considerably altered. We first analyzed in detail the total gamma ray spectrum from gravitino decays, in particular the dependence of the new continuum contribution on gaugino and gravitino masses. Then, we computed the expected gamma ray flux from gravitino decays and derive new bounds on $\xi_\tau$ and on the gravitino lifetime from recent FERMI data.  The gravitino lifetime was shown to be constrained by such data to be larger than about $10^{27}$ s to $10^{29}$ s, depending on the gravitino and gaugino masses. We also investigated the indirect detection of gravitinos via antimatter signals.  In the range of gravitino masses we consider, they are entirely due to the three-body final states and had not been studied before. After analyzing the positron and antiproton spectra from gravitino decay and obtaining the expected fluxes at earth, we demonstrated  that only the antiproton signal can be significant enough to be of interest  for future experiments.

\acknowledgments
We would like to thank A. Ibarra and C. Weniger for pointing out an error in equation (\ref{eq:gflux2}) that led, in the previous version of this paper, to an overestimation of the continuum flux. Our main results are not significantly modified due to this correction. K.Y. Choi was partly supported by the Korea Research
Foundation Grant funded by the Korean Government (KRF-2008-341-C00008)
and by the second stage of Brain Korea 21 Project in 2006.
C. E. Y. is supported by the \emph{Juan de la Cierva} program of the MICINN of Spain. He acknowledges additional support from the MICINN Consolider-Ingenio 2010 Programme under grant MULTIDARK CSD2009-00064, from the MCIINN under Proyecto Nacional FPA2009-08958, and from the CAM under grant HEPHACOS S2009/ESP-1473. D.R was partly supported by Sostenibilidad-UdeA/2009 grant. O. Z. acknowledges financial support to \textit{Direcci\'on de Investigaci\'on y Proyectos} of \textit{Escuela de Ingenier\'ia de Antioquia}.

\appendix
\section{Analytical formulas for the three-body decays}
The gravitinos can decay to $f+\bar{f}$ and neutrino mediated by virtual
photon or $Z$-boson, or $f+\bar{f}'$ and chargeed lepton by virtual $W$-boson.
Here we summarise the main formulas for the gravitino three body decays.

The differential decay rate for the three body-decay process is given
by~\footnote{The analytic expression of the three-body decay are also shown 
in~\cite{Feng:2004mt},~\cite{Kohri:2005wn} and~\cite{Cyburt:2009pg}
for R-parity conserving case.}
\dis{
\frac{d\Gamma}{dsdt}=\frac{N_c}{256\pi^3\mgravitino^3}\overline{|{\mathcal M}|^2},
}
where $N_c$ is the color factor (3 for $q$ $\bar{q}$ final states and 1 for
lepton pairs.
The parameters $s$ and $t$ are the invariant masses of the $f\bar{f}$ (or
$f\bar{f}'$) and $f\nu$ systems respectively.
We can define the masses and  momenta of the relevant fields as
($\mgravitino, p=k_1+k_2+q$) for Gravitino($\psi_\mu$), ($m_1,k_1$) for a fermion ($f$),
 ($m_2,k_2$) for the other fermion ($\bar{f}$ or $\bar{f}'$), ($m_3, q$)
for the neutrino or charged lepton and ($m_V,k=k_1+k_2$) for an 
intermediate vector boson.
Then the invariant masses $s$ and $t$ are
\dis{
s=(k_1+k_2)^2,\qquad t=(k_1+q)^2.
}
The range of these parameters are

\dis{
(m_1+m_2)^2 \leq s\leq(\mgravitino-m_3)^2,
}
and for fixed $s$, $t$ are in the range
\dis{
t_{min}\leq t\leq t_{max},
}
where
\dis{
t_{max,min}=m_3^2+m_1^2+\frac{1}{2s}\left[(\mgravitino^2-s-m_3^2)(s-m_2^2+m_1^2)\pm \lambda^{1/2}(s,\mgravitino^2,m_3^2)\lambda^{1/2}(s,m_1^2,m_2^2) \right].
}
The function $\lambda(x,y,z)$ is defined as
\dis{
\lambda(x,y,z)=x^2+y^2+z^2-2xy-2yz-2zx.
}
%%%%%%%%%%%%%%
\subsection{$\gravitino \to \gamma^*/Z^* + \nu \to f+\bar{f} +\nu$}
The photon and/or $Z$-boson mediated three-body decay rate amplitude 
can be divided into three parts: photon-mediated, $Z$-boson mediated and the interference of the two,
\dis{
i{\mathcal M}=i{\mathcal M}_\gamma+i{\mathcal M}_Z+i{\mathcal M}_{int}.
}
For the range of the gravitino masses we consider, the photon-mediated diagram is subdominant and the amplitude is well approximated by the $Z$-mediated part. The two diagrams that contribute to gravitino decay into $Z^*\nu_\tau$ ($\to \sum_ff\bar f\nu_\tau$) are shown in figure \ref{fig:Zdiag}.
\begin{center}
\FIGURE{\begin{tabular}{ccc}\includegraphics[scale=0.3]{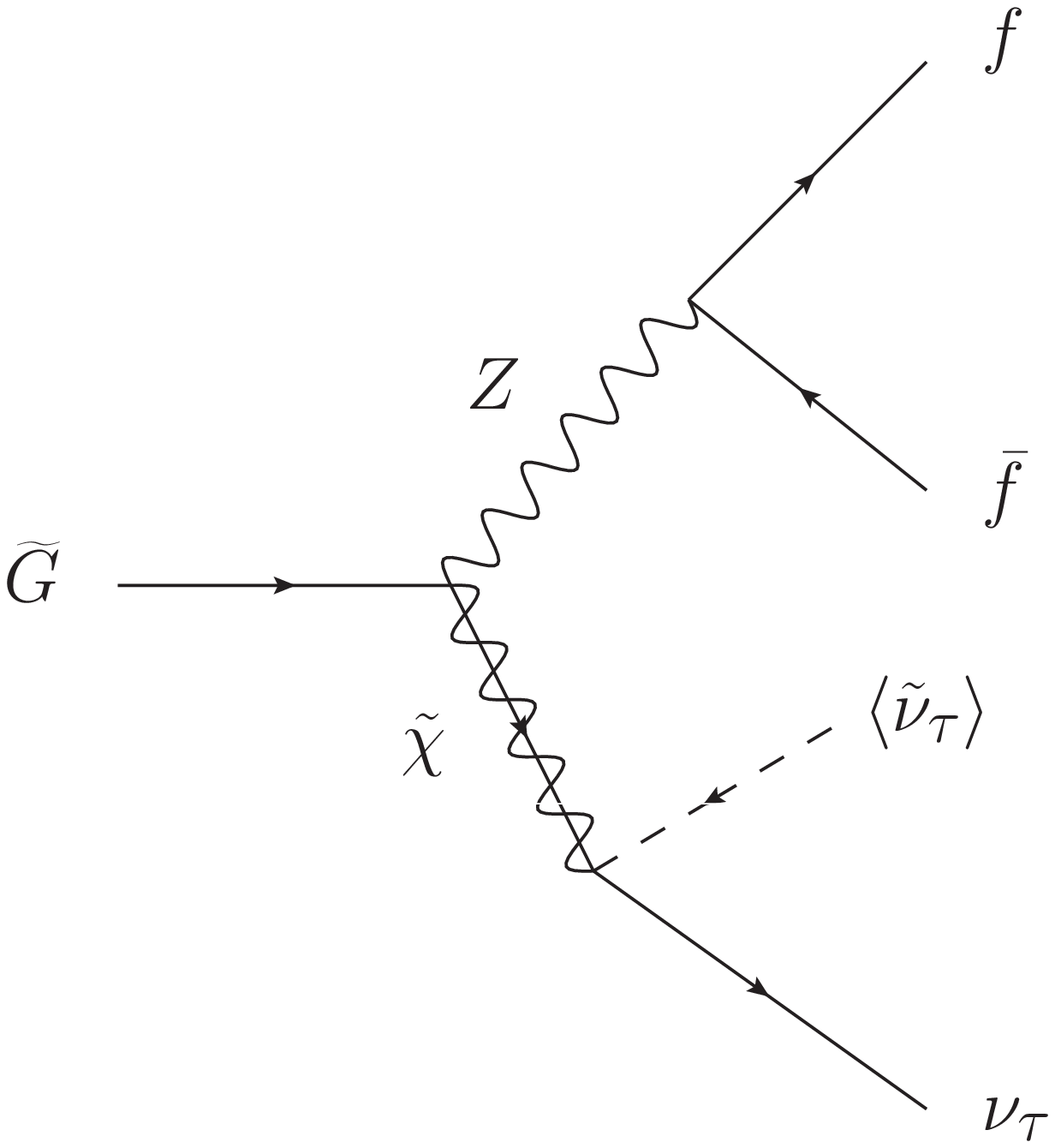} & \hspace{1cm} & \includegraphics[scale=0.34]{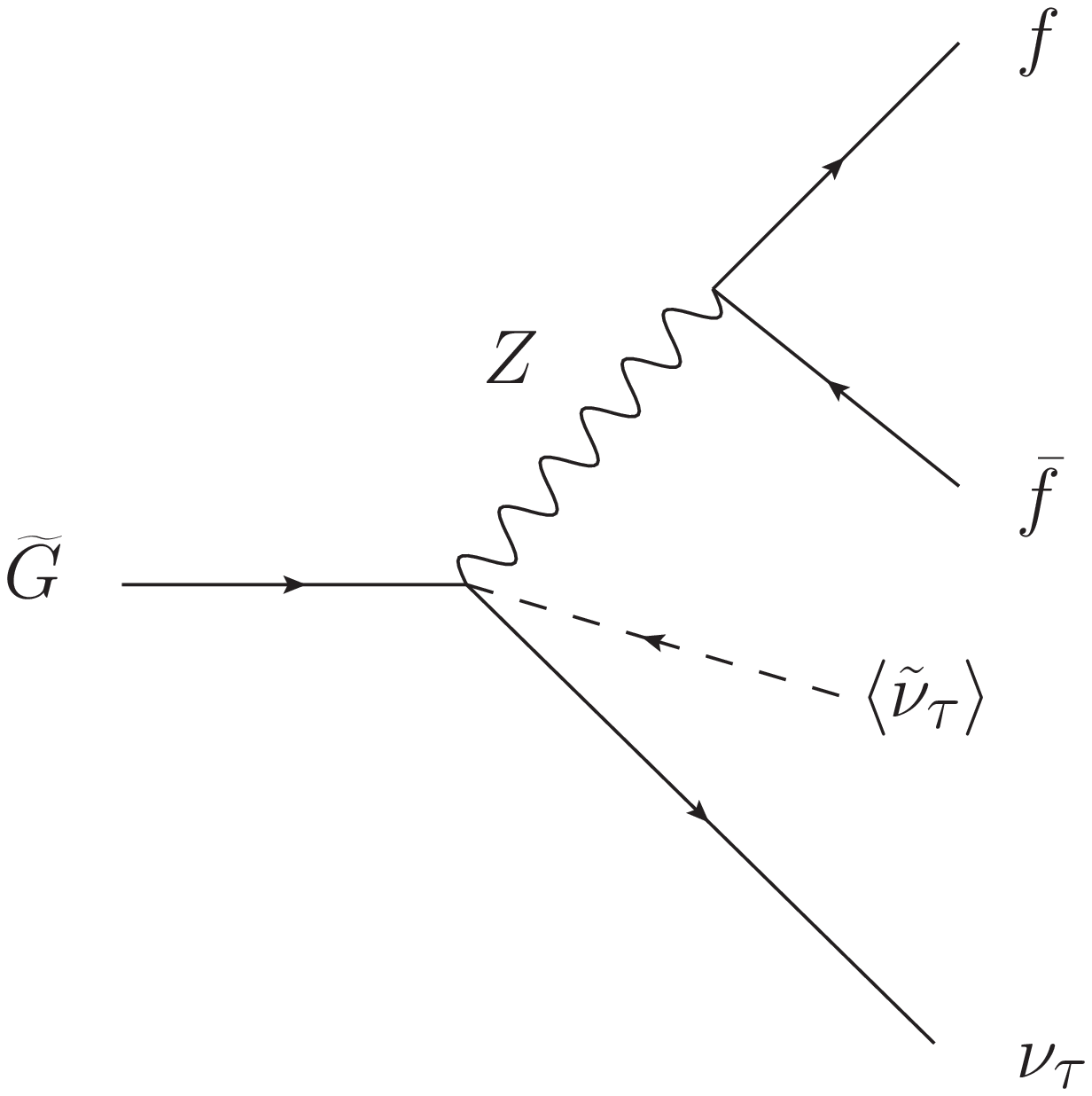}\end{tabular}\caption{The two diagrams that contribute to gravitino decay into $Z^*\nu_\tau$}\label{fig:Zdiag}}
\end{center}
The evaluation of those diagrams yields 
\dis{
\overline {|{\mathcal M}_Z|^2}=\frac{g^2\xi_\tau^2}{64 Mp^2 \cos^2\theta_W}\frac{2}{3 {\mgravitino}^3 {\mz}^4 \left[{\Gamma_Z}^2
   {\mz}^2+\left({\mz}^2-s\right)^2\right]}TrZ0,
}
where $g$ is $SU(2)$ gauge coupling, $\theta_W$ is weak mixing angle
and $\Gamma_Z$ is the decay width of $Z$-boson.
The trace part is given by, in terms of the invariant masses $s$ and $t$ and
neglecting the masses of the final particles,  
\dis{
&TrZ0\\
&=\left(30 {\ca}^2 s {\mz}^6+30 {\cv}^2 s {\mz}^6-36 {\ca}
   {\cv} s {\mz}^6+24 {\ca}^2 t {\mz}^6+24 {\cv}^2 t
   {\mz}^6\right) {\mgravitino}^5\\
   & \quad
   +\left(-30 {\ca}^2 s^2 {\mz}^6-30
   {\cv}^2 s^2 {\mz}^6+36 {\ca} {\cv} s^2 {\mz}^6-24
   {\ca}^2 t^2 {\mz}^6-24 {\cv}^2 t^2 {\mz}^6\right.\\
   & \qquad\left.-48 {\ca}^2
   s t {\mz}^6-48 {\cv}^2 s t {\mz}^6+72 {\ca} {\cv} s t
   {\mz}^6\right) {\mgravitino}^3\\
&\quad+\left(24 {\ca}^2 s t^2 {\mz}^6+24
   {\cv}^2 s t^2 {\mz}^6+24 {\ca}^2 s^2 t {\mz}^6+24
   {\cv}^2 s^2 t {\mz}^6\right) {\mgravitino}\\
%%%%
   &\quad+{\Uz} \left[\left(-38
   {\ca}^2 s {\mz}^5-38 {\cv}^2 s {\mz}^5+76 {\ca}
   {\cv} s {\mz}^5\right) {\mgravitino}^6\right.\\
   & \qquad\left.+\left(22 {\ca}^2 s^2
   {\mz}^5+22 {\cv}^2 s^2 {\mz}^5-92 {\ca} {\cv} s^2
   {\mz}^5+14 {\ca}^2 s t {\mz}^5+14 {\cv}^2 s t
   {\mz}^5\right.\right.\\
   & \qquad\quad\left.\left.-152 {\ca} {\cv} s t {\mz}^5\right)
   {\mgravitino}^4\right.\\
   & \qquad\left.+\left(16 {\ca}^2 s^3 {\mz}^5+16 {\cv}^2 s^3
   {\mz}^5+16 {\ca} {\cv} s^3 {\mz}^5-14 {\ca}^2 s t^2
   {\mz}^5-14 {\cv}^2 s t^2 {\mz}^5\right.\right.\\
   & \qquad\quad\left.\left.-10 {\ca}^2 s^2 t
   {\mz}^5-10 {\cv}^2 s^2 t {\mz}^5+32 {\ca} {\cv} s^2 t
   {\mz}^5\right) {\mgravitino}^2\right.\\
   & \qquad\left.-4 {\ca}^2 {\mz}^5 s^2 t^2-4
   {\cv}^2 {\mz}^5 s^2 t^2-4 {\ca}^2 {\mz}^5 s^3 t-4
   {\cv}^2 {\mz}^5 s^3 t\right]\\
&\quad+{\Uz}^2 \left[\left(14
   {\ca}^2 s {\mz}^4+14 {\cv}^2 s {\mz}^4-28 {\ca}
   {\cv} s {\mz}^4\right) {\mgravitino}^7\right.\\
   & \qquad\left.+\left(-10 {\ca}^2 s^2
   {\mz}^4-10 {\cv}^2 s^2 {\mz}^4+20 {\ca} {\cv} s^2
   {\mz}^4-26 {\ca}^2 s t {\mz}^4-26 {\cv}^2 s t
   {\mz}^4\right.\right.\\
   & \qquad\quad\left.\left.+56 {\ca} {\cv} s t {\mz}^4\right)
   {\mgravitino}^5\right.\\
   & \qquad\left.+\left(2 {\ca}^2 s^3 {\mz}^4+2 {\cv}^2 s^3
   {\mz}^4+20 {\ca} {\cv} s^3 {\mz}^4+26 {\ca}^2 s t^2
   {\mz}^4+26 {\cv}^2 s t^2 {\mz}^4\right.\right.\\
   & \qquad\quad\left.\left.+34 {\ca}^2 s^2 t
   {\mz}^4+34 {\cv}^2 s^2 t {\mz}^4+16 {\ca} {\cv} s^2 t
   {\mz}^4\right) {\mgravitino}^3\right.\\
   & \qquad\left.+\left(-6 {\ca}^2 s^4 {\mz}^4-6
   {\cv}^2 s^4 {\mz}^4-12 {\ca} {\cv} s^4 {\mz}^4-8
   {\ca}^2 s^2 t^2 {\mz}^4-8 {\cv}^2 s^2 t^2 {\mz}^4\right.\right.\\
   & \qquad\quad\left.\left.-8
   {\ca}^2 s^3 t {\mz}^4-8 {\cv}^2 s^3 t {\mz}^4-24
   {\ca} {\cv} s^3 t {\mz}^4\right] {\mgravitino}\right).
}
Here $\Uz$ is the zino-zino mixing parameter (see e.g. \cite{Covi:2008jy} for its definition), which we assumed to be real. The coefficients $\cv$ and $\ca$ are defined as
\dis{
\cv=\frac12 T_3(f) -Q(f)\sin^2\theta_W,\qquad \ca=-\frac12T_3(f),
}
where $T_3$ and $Q$ are the third component of weak isospin and electric
charge, respectively and we used $P_L=(1-\gamma_5)/2$. 
For example, for the left-handed neutrino, $\cv=1/4$ and $\ca=-1/4$.

%%%%%%%%%%%%
\subsection{$\gravitino \to W^* + \tau^- \to f+\bar{f}'+\tau^- $}
The two diagrams that contribute to the decay of the gravitino into $W^*\tau$ ($\sum_f f\bar f'\tau$) are shown in figure \ref{fig:Wdiag}. The resulting squared amplitude for that process is given by
\dis{
\overline {|{\mathcal M}_W|^2}=\frac{g^2\xi^2}{64 Mp^2}\frac{2}{3 \mgravitino^3 \mw^4 \left[\Gamma_W^2
   \mw^2+\left(\mw^2-s\right)^2\right]}TrW0,
}
where $g$ is $SU(2)$ gauge coupling and $\Gamma_W$ is the decay width of $W$-boson.

The trace part is given by, neglecting the masses of final particles, 
\dis{
&TrW0\\
&=(96 s \mw^6+48 t \mw^6) \mgravitino^5+(-96 s^2 \mw^6-48 t^2 \mw^6-168 s t \mw^6) \
\mgravitino^3\\
&\quad +(48 s t^2 \mw^6+48 s^2 t \mw^6) \mgravitino\\
&\quad +\Uw \left[-152 \mw^5 s \mgravitino^6+(136 s^2 \
\mw^5+180 s t \mw^5) \mgravitino^4\right.\\
   & \qquad\left.+(16 s^3 \mw^5-28 s t^2 \mw^5-52 s^2 t \mw^5) \
\mgravitino^2-8 \mw^5 s^2 t^2-8 \mw^5 s^3 t\right]\\
&\quad+\Uw^2 \left[56 \mw^4 s \mgravitino^7+(-40 s^2 \
\mw^4-108 s t \mw^4) \mgravitino^5\right.\\
   & \qquad\left.+(-16 s^3 \mw^4+52 s t^2 \mw^4+52 s^2 t \mw^4) \
\mgravitino^3+(8 \mw^4 s^3 t-16 \mw^4 s^2 t^2) \mgravitino\right],
}
where $U_{\wino\wino}$, the wino-wino mixing parameter (see e.g. \cite{Covi:2008jy}),  was assumed to be real.


\begin{thebibliography}{99}

\bibitem{Dunkley:2008ie}
  J.~Dunkley {\it et al.}  [WMAP Collaboration],
  %``Five-Year Wilkinson Microwave Anisotropy Probe (WMAP) Observations:
  %Likelihoods and Parameters from the WMAP data,''
  Astrophys.\ J.\ Suppl.\  {\bf 180} (2009) 306
  [arXiv:0803.0586 [astro-ph]].
  %%CITATION = APJSA,180,306;%%


\bibitem{Takayama:2000uz}
F.~Takayama and M.~Yamaguchi, 
%``Gravitino dark matter without R-parity,'' 
Phys.\ Lett.\ B {\bf 485} (2000) 388 [arXiv:hep-ph/0005214]. 
%%CITATION = PHLTA,B485,388;%%


\bibitem{Buchmuller:2007ui}
W.~Buchmuller, L.~Covi, K.~Hamaguchi, A.~Ibarra and T.~Yanagida, 
%``Gravitino dark matter in R-parity breaking vacua,'' 
JHEP {\bf 0703} (2007) 037 [arXiv:hep-ph/0702184]. 
%%CITATION = JHEPA,0703,037;%%


\bibitem{Bertone:2007aw}
  G.~Bertone, W.~Buchmuller, L.~Covi and A.~Ibarra,
  %``Gamma-Rays from Decaying Dark Matter,''
  JCAP {\bf 0711} (2007) 003
  [arXiv:0709.2299 [astro-ph]].
  %%CITATION = JCAPA,0711,003;%%

\bibitem{Ishiwata:2008cu}
  K.~Ishiwata, S.~Matsumoto and T.~Moroi,
  %``High Energy Cosmic Rays from the Decay of Gravitino Dark Matter,''
  Phys.\ Rev.\  D {\bf 78} (2008) 063505
  [arXiv:0805.1133 [hep-ph]].
  %%CITATION = PHRVA,D78,063505;%%

\bibitem{Ibarra:2007wg}
  A.~Ibarra and D.~Tran,
  %``Gamma Ray Spectrum from Gravitino Dark Matter Decay,''
  Phys.\ Rev.\ Lett.\  {\bf 100} (2008) 061301
  [arXiv:0709.4593 [astro-ph]].
  %%CITATION = PRLTA,100,061301;%%



\bibitem{Covi:2008jy}
  L.~Covi, M.~Grefe, A.~Ibarra and D.~Tran,
  %``Unstable Gravitino Dark Matter and Neutrino Flux,''
  JCAP {\bf 0901} (2009) 029
  [arXiv:0809.5030 [hep-ph]].
  %%CITATION = JCAPA,0901,029;%%
\bibitem{Ibarra:2008qg}
 A.~Ibarra and D.~Tran,
 %``Antimatter Signatures of Gravitino Dark Matter Decay,''
  JCAP {\bf 0807} (2008) 002 
%[arXiv:0804.4596 [astro-ph]].
  %%CITATION = JCAPA,0807,002;%%

\bibitem{Choi:2009ng}
 K.~Y.~Choi, D.~E.~Lopez-Fogliani, C.~Munoz and R.~R.~de Austri,
  %``Gamma-ray detection from gravitino dark matter decay in the $\mu\nu$SSM,''
  JCAP {\bf 1003} (2010) 028
  [arXiv:0906.3681 [hep-ph]].
  %%CITATION = JCAPA,1003,028;%%
A.~A.~Abdo {\it et al.},
  %``Fermi LAT Search for Photon Lines from 30 to 200 GeV and Dark Matter
  %Implications,''
  Phys.\ Rev.\ Lett.\  {\bf 104} (2010) 091302
  [arXiv:1001.4836 [astro-ph.HE]].
  %%CITATION = PRLTA,104,091302;%%

%\cite{Choi:2010xn}
\bibitem{Choi:2010xn}
  K.~Y.~Choi and C.~E.~Yaguna,
  %``New decay modes of gravitino dark matter,''
  Phys.\ Rev.\  D {\bf 82} (2010) 015008
  [arXiv:1003.3401 [hep-ph]].
  %%CITATION = PHRVA,D82,015008;%%


\bibitem{Bagger:1990qh}
  J.~Bagger and J.~Wess,
  ``Supersymmetry and supergravity''.
  %%CITATION = JHU-TIPAC-9009;%%

\bibitem{Grefe:2008zz}
  M.~Grefe, DESY-THESIS-2008-043.
  %``Neutrino signals from gravitino dark matter with broken R-parity,''
  %%CITATION = DESY-THESIS-2008-043;%%

\bibitem{Sjostrand:2006za}
  T.~Sjostrand, S.~Mrenna and P.~Z.~Skands,
  %``PYTHIA 6.4 Physics and Manual,''
  JHEP {\bf 0605}, 026 (2006)
  [arXiv:hep-ph/0603175].
  %%CITATION = JHEPA,0605,026;%%

\bibitem{Buchmuller:2009xv}
  W.~Buchmuller, A.~Ibarra, T.~Shindou, F.~Takayama and D.~Tran,
  %``Probing Gravitino Dark Matter,''
  JCAP {\bf 0909} (2009) 021
%  [arXiv:0906.1187 [hep-ph]].
  %%CITATION = JCAPA,0909,021;%%


\bibitem{Navarro:1996gj}
  J.~F.~Navarro, C.~S.~Frenk and S.~D.~M.~White,
  %``A Universal Density Profile from Hierarchical Clustering,''
  Astrophys.\ J.\  {\bf 490}, 493 (1997)
  [arXiv:astro-ph/9611107].
  %%CITATION = ASJOA,490,493;%%

%\cite{Catena:2009mf}
\bibitem{Catena:2009mf}
  R.~Catena and P.~Ullio,
  %``A novel determination of the local dark matter density,''
  JCAP {\bf 1008} (2010) 004
  [arXiv:0907.0018 [astro-ph.CO]].
  %%CITATION = JCAPA,1008,004;%%


%\cite{Abdo:2010nc}
\bibitem{Abdo:2010nc}
  A.~A.~Abdo {\it et al.},
  %``Fermi LAT Search for Photon Lines from 30 to 200 GeV and Dark Matter
  %Implications,''
  Phys.\ Rev.\ Lett.\  {\bf 104} (2010) 091302
  [arXiv:1001.4836 [astro-ph.HE]].
  %%CITATION = PRLTA,104,091302;%%
\bibitem{Abdo:2010nz}
  A.~A.~Abdo {\it et al.}  [The Fermi-LAT collaboration],
  %``The Spectrum of the Isotropic Diffuse Gamma-Ray Emission Derived From
  %First-Year Fermi Large Area Telescope Data,''
  Phys.\ Rev.\ Lett.\  {\bf 104}, 101101 (2010)
  [arXiv:1002.3603 [astro-ph.HE]].
  %%CITATION = PRLTA,104,101101;%%

\bibitem{otherconstraints}
M.~Cirelli, P.~Panci and P.~D.~Serpico,
  %``Diffuse gamma ray constraints on annihilating or decaying Dark Matter after
  %Fermi,''
  arXiv:0912.0663 [astro-ph.CO].
  %%CITATION = ARXIV:0912.0663;%%

G.~Hutsi, A.~Hektor and M.~Raidal,
  %``Implications of the Fermi-LAT diffuse gamma-ray measurements on
  %annihilating or decaying Dark Matter,''
  arXiv:1004.2036 [astro-ph.HE].
  %%CITATION = ARXIV:1004.2036;%%
A.~A.~Abdo {\it et al.}  [Fermi-LAT Collaboration],
  %``Constraints on Cosmological Dark Matter Annihilation from the Fermi-LAT
  %Isotropic Diffuse Gamma-Ray Measurement,''
  JCAP {\bf 1004} (2010) 014
  [arXiv:1002.4415 [astro-ph.CO]].
  %%CITATION = JCAPA,1004,014;%%
K.~N.~Abazajian, P.~Agrawal, Z.~Chacko and C.~Kilic,
  %``Conservative Constraints on Dark Matter from the Fermi-LAT Isotropic
  %Diffuse Gamma-Ray Background Spectrum,''
  arXiv:1002.3820 [astro-ph.HE].
  %%CITATION = ARXIV:1002.3820;%%


%\cite{Delahaye:2007fr}
\bibitem{Delahaye:2007fr}
  T.~Delahaye, R.~Lineros, F.~Donato, N.~Fornengo and P.~Salati,
  %``Positrons from dark matter annihilation in the galactic halo: theoretical
  %uncertainties,''
  Phys.\ Rev.\  D {\bf 77} (2008) 063527
  [arXiv:0712.2312 [astro-ph]].
  %%CITATION = PHRVA,D77,063527;%%
\bibitem{Cirelli:2008id}
  M.~Cirelli, R.~Franceschini and A.~Strumia,
  %``Minimal Dark Matter predictions for galactic positrons, anti-protons,
  %photons,''
  Nucl.\ Phys.\  B {\bf 800}, 204 (2008)
  [arXiv:0802.3378 [hep-ph]].
  %%CITATION = NUPHA,B800,204;%%
   E.~Nardi, F.~Sannino and A.~Strumia,
  %``Decaying Dark Matter can explain the electron/positron excesses,''
  JCAP {\bf 0901} (2009) 043
  [arXiv:0811.4153 [hep-ph]].
  %%CITATION = JCAPA,0901,043;%%



\bibitem{Ibarra:2009dr}
  A.~Ibarra, D.~Tran and C.~Weniger,
  %``Decaying Dark Matter in Light of the PAMELA and Fermi LAT Data,''
  JCAP {\bf 1001}, 009 (2010)
  [arXiv:0906.1571 [hep-ph]].
  %%CITATION = JCAPA,1001,009;%%



%\cite{Grasso:2009ma}
\bibitem{Grasso:2009ma}
  D.~Grasso {\it et al.}  [FERMI-LAT Collaboration],
  %``On possible interpretations of the high energy electron-positron spectrum
  %measured by the Fermi Large Area Telescope,''
  Astropart.\ Phys.\  {\bf 32} (2009) 140
  [arXiv:0905.0636 [astro-ph.HE]].
  %%CITATION = APHYE,32,140;%%

%\cite{Gleeson:1968zz}
\bibitem{Gleeson:1968zz}
  L.~J.~Gleeson and W.~I.~Axford,
  %``Cosmic Rays in the Interplanetary Medium,''
  Astrophys.\ J.\  {\bf 149} (1967) L115.
  %``Solar Modulation of Galactic Cosmic Rays,''
  Astrophys.\ J.\  {\bf 154} (1968) 1011.
  %%CITATION = ASJOA,154,1011;%%


%\cite{Adriani:2008zr}
\bibitem{Adriani:2008zr}
  O.~Adriani {\it et al.}  [PAMELA Collaboration],
  %``An anomalous positron abundance in cosmic rays with energies 1.5-100 GeV,''
  Nature {\bf 458} (2009) 607
  [arXiv:0810.4995 [astro-ph]].
  %%CITATION = NATUA,458,607;%%



%\cite{Donato:2003xg}
\bibitem{Donato:2003xg}
  F.~Donato, N.~Fornengo, D.~Maurin and P.~Salati,
  %``Antiprotons in cosmic rays from neutralino annihilation,''
  Phys.\ Rev.\  D {\bf 69} (2004) 063501
  [arXiv:astro-ph/0306207].
  %%CITATION = PHRVA,D69,063501;%%
  
%\cite{Bringmann:2006im}
\bibitem{Bringmann:2006im}
  T.~Bringmann and P.~Salati,
  %``The galactic antiproton spectrum at high energies: background   expectation
  %vs. exotic contributions,''
  Phys.\ Rev.\  D {\bf 75} (2007) 083006
  [arXiv:astro-ph/0612514].
  %%CITATION = PHRVA,D75,083006;%%

\bibitem{Bai:2009ka}
  Y.~Bai, M.~Carena and J.~Lykken,
  %``The PAMELA excess from neutralino annihilation in the NMSSM,''
  Phys.\ Rev.\  D {\bf 80}, 055004 (2009)
  [arXiv:0905.2964 [hep-ph]].
  %%CITATION = PHRVA,D80,055004;%%

%\cite{Adriani:2008zq}
\bibitem{Adriani:2008zq}
  O.~Adriani {\it et al.},
  %``A new measurement of the antiproton-to-proton flux ratio up to 100 GeV in
  %the cosmic radiation,''
  Phys.\ Rev.\ Lett.\  {\bf 102} (2009) 051101
  [arXiv:0810.4994 [astro-ph]].
  %%CITATION = PRLTA,102,051101;%%
{\it et al.}  [PAMELA Collaboration],
  %``PAMELA results on the cosmic-ray antiproton flux from 60 MeV to 180 GeV in
  %kinetic energy,''
  arXiv:1007.0821 [astro-ph.HE].
  %%CITATION = ARXIV:1007.0821;%%



%\cite{Feng:2004mt}
\bibitem{Feng:2004mt}
  J.~L.~Feng, S.~Su and F.~Takayama,
  %``Supergravity with a gravitino LSP,''
  Phys.\ Rev.\  D {\bf 70} (2004) 075019
  [arXiv:hep-ph/0404231].
  %%CITATION = PHRVA,D70,075019;%%

%\cite{Kohri:2005wn}
\bibitem{Kohri:2005wn}
  K.~Kohri, T.~Moroi and A.~Yotsuyanagi,
  %``Big-bang nucleosynthesis with unstable gravitino and upper bound on the
  %reheating temperature,''
  Phys.\ Rev.\  D {\bf 73} (2006) 123511
  [arXiv:hep-ph/0507245].
  %%CITATION = PHRVA,D73,123511;%%

%\cite{Cyburt:2009pg}
\bibitem{Cyburt:2009pg}
  R.~H.~Cyburt, J.~Ellis, B.~D.~Fields, F.~Luo, K.~A.~Olive and V.~C.~Spanos,
  %``Nucleosynthesis Constraints on a Massive Gravitino in Neutralino Dark
  %Matter Scenarios,''
  JCAP {\bf 0910} (2009) 021
  [arXiv:0907.5003 [astro-ph.CO]].
  %%CITATION = JCAPA,0910,021;%%

%\cite{newglines}
\bibitem{newglines}
  G. Vertongen, C. Weniger,
  %``Hunting 1-500 GeV Dark Matter Gamma-Ray Lines with the Fermi LAT ,''  
  [arXiv:1101.2610].



\end{thebibliography}
\end{document}